# Tuning interactions between spins in a superconductor

Hao Ding[1†], Yuwen Hu[1†], Mallika T. Randeria[1‡], Silas Hoffman[2,3], Oindrila Deb[2], Jelena Klinovaja[2], Daniel Loss[2], Ali Yazdani[1]*

[1]Joseph Henry Laboratories & Department of Physics, Princeton University, Princeton, NJ 08544, USA.
[2]Department of Physics, University of Basel, Klingelbergstrasse 82, CH-4056 Basel, Switzerland.
[3]Department of Physics, University of Florida, Gainesville, FL 32611, USA.

*To whom correspondence may be addressed. Email: yazdani@princeton.edu
[†]These authors contributed equally to this work.
[‡]Present address: Department of Physics, Massachusetts Institute of Technology, Cambridge, Massachusetts 02139, USA.

**This PDF file includes:**

Main Text
Materials and Methods
References (1–48)
Figures 1 to 4

Figures S1 to S27
Supplementary Text (Sections I to XIV)
Tables S1 and S2
SI References (1–9)

Abstract

Novel many-body and topological electronic phases can be created in assemblies of interacting spins coupled to a superconductor, such as one-dimensional topological superconductors with Majorana zero modes (MZMs) at their ends. Understanding and controlling interactions between spins and the emergent band structure of the in-gap Yu-Shiba-Rusinov (YSR) states they induce in a superconductor are fundamental for engineering such phases. Here, by precisely positioning magnetic adatoms with a scanning tunneling microscope (STM), we demonstrate both the tunability of exchange interaction between spins and precise control of the hybridization of YSR states they induce on the surface of a bismuth (Bi) thin film that is made superconducting with the proximity effect. In this platform, depending on the separation of spins, the interplay between Ruderman-Kittel-Kasuya-Yosida (RKKY) interaction, spin-orbit coupling, and surface magnetic anisotropy stabilizes different types of spin alignments. Using high-resolution STM spectroscopy at millikelvin temperatures, we probe these spin alignments through monitoring the spin-induced YSR states and their energy splitting. Such measurements also reveal a quantum phase transition between the ground states with different electron number parity for a pair of spins in a superconductor tuned by their separation. Experiments on larger assemblies show that spin-spin interactions can be mediated in a superconductor over long distances. Our results show that controlling hybridization of the YSR states in this platform provides the possibility of engineering the band structure of such states for creating topological phases.

**Significance**

Majorana zero modes (MZMs) have been proposed as the building blocks for fault-tolerant topological quantum computation. Recent experiments have found both spin and charge signatures of MZMs at the ends of spin chains on superconducting surfaces. However, the properties of such chains have not been reliably controlled experimentally. Here we present a platform in which long-range spin-spin interactions and hybridization of spin-induced in-gap states on the surface of a superconductor can be tuned with unprecedented precision by changing the distance between the spins using atomic manipulation technique. This capability is required for tailoring magnetic textures and engineering in-gap states band structures in spin assemblies and opens up the possibility of exploring new topological superconducting phases with tunable properties.

**Main Text**

**Introduction**

The goal of realizing topological electronic phases using combinations of superconductivity, magnetism, and spin-orbit interaction has motivated efforts in creating spin chains and other magnetic assemblies on the surfaces of superconductors (1–10). There is now considerable evidence that a topological superconducting phase forms in closely-packed one-dimensional ferromagnetic chains made of magnetic atoms on the surface of a superconductor with strong spin-orbit coupling, where at the ends of the chains Majorana zero modes (MZMs) have been detected in various scanning tunneling microscope (STM) experiments (11–15). The next step in advancing the study of MZMs in atomic chains is to build chains using STM atomic manipulation techniques starting from single magnetic atoms to be able to systematically probe the topological phase diagram (1, 3, 5–8). These experiments could also make it possible to test the MZMs' non-Abelian properties, such as their fusion rules, and perhaps ultimately to braid them (16). The key parameters of the topological phase diagram of atomic chains are their spin texture and the bandwidth of their overlapping in-gap Yu-Shiba-Rusinov (YSR) states. Demonstration of atomic manipulation experiments that can control these parameters would open up a wide range of future experiments. While there have been efforts to successfully create closely-packed iron chains with nearest neighbor exchange interaction on the surface of superconducting rhenium using STM atomic manipulation (17), a platform which meets the above tunability requirements has not been realized.

Here we show that magnetic atoms on the surface of bismuth (Bi) thin films made superconducting by the proximity effect provides such a platform. The relatively large Fermi wavelength of this surface as compared to its atomic lattice spacing, and its strong spin-orbit interaction make it possible to use atomic manipulation with STM to create spin assemblies with different spin alignments. By fine tuning the distance between pairs of interacting spins, we have precisely measured the splitting of their YSR states and observed a quantum phase transition between phases with different electron number parity tuned by their separation. In larger spin assemblies, we find evidence that non-collinear alignment of spins controls the splitting of YSR states, thereby illustrating the potential for YSR band structure engineering with topological properties in this platform.

**Results and Discussion**

Experimental Platform

To realize our platform, we grow epitaxial Bi ultra-thin films (6 monolayers) on the surface of superconducting Nb(110) substrate *in situ* (see Materials and Methods and Fig. S1) in a homebuilt millikelvin ultra-high vacuum STM system (18). The proximity-induced superconducting gap on the surface of the Bi thin film ($\Delta$) is 1.52 meV, as large as that measured on the bare Nb surface (Fig. S1C). The Bi films have a Bi(110) like zig-zag atomic structure (Fig. S1B) but are more likely in a metastable pseudocubic (PC) phase, as previous studies find to be most stable at such a thickness (19, 20). For either PC phase (19, 20) or Bi(110) surface (21), the largest Fermi pocket is near $\Gamma$ point at the Brillouin zone center with Fermi wavelength ($\lambda_F$) ~ 15 Å, which is much shorter than the proximitized Bi in-plane coherence length ($\xi$ ~ 600 Å), making Ruderman-Kittel-Kasuya-Yosida (RKKY) interaction one of the dominant interactions between spins when their separation is at the scale of $\lambda_F$ on the surface (22, 23). More importantly, the $\lambda_F$ of Bi surface is almost three times as long as that of Nb (5.4 Å) and more than four times that of the smallest atomic spacing (3.30 Å) on the Bi surface with zig-zag lattice (Fig. S1B), which allows for significant tunability of the RKKY interaction by placing spins with separation smaller than $\lambda_F$ on the Bi lattice. Therefore, this surface provides an ideal platform to tune interactions between spins in the presence of superconductivity. Typically, spin-polarized STM measurements are used to detect components of the spin polarization of magnetic atoms on surfaces; such measurements are most reliable when used in combination with the application of a magnetic field (24). However, an applied magnetic field can not only modify the magnetic configuration of atoms on a surface but also suppress superconductivity. Here we probe magnetic interactions between spins in a superconductor by detecting the YSR states that magnetic impurities induce within the host superconductor energy gap. In the simplest case of a single magnetic impurity, a single YSR state appears as a pair of electron- and hole-like partner peaks in the local density of states (LDOS) at energies $E = \pm\Omega$ ($|\Omega| \leq \Delta$) with respect to the chemical potential (25–29). For weak exchange coupling between the impurity and the superconductor, YSR states lie near the gap edge at $E = \pm\Delta$ and move closer to the middle of the gap with increasing exchange interaction. Two spins induce two YSR states that can spatially overlap and hybridize, giving rise to the formation of bonding and anti-bonding YSR states and the observation of four (two pairs) peaks in LDOS with their energy splitting proportional to the overlap between the YSR wavefunctions (23, 29–37). For distances much smaller than the coherence length ($r \ll \xi$), the magnetic configuration of the spins has a rich dependence on the RKKY interaction, direct exchange interaction, and overlap of YSR states themselves (22, 23, 34). Nonetheless, the shift and splitting of YSR states can be used to probe the magnetic ground state as well as the interactions between spins in a superconductor (23, 29–37).

We probe the interactions between spins in the superconducting state by constructing precisely positioned pairs of magnetic atoms on the proximitized Bi thin film surface and probe their properties with STM spectroscopy at millikelvin temperatures (with energy resolution of 100 µeV, see Materials and Methods and Fig. S2). For this study, we use gadolinium (Gd) atoms as magnetic impurities ($S = 7/2$), which are evaporated on the pre-cooled Bi surface *in situ* in the microscope prior to measurements at the lowest temperature. Figure 1 A–D display STM topographs of four examples of such STM-constructed Gd pairs. Their spectroscopic characteristics in differential conductance (*dI/dV*) measurements performed on one of the atoms in each pair are shown in Fig. 1 E–H. These measurements show either two (at energies $E = \pm\Omega$) or four peaks (at $E = \pm\Omega_1, \pm\Omega_2$) in LDOS appearing inside the superconducting gap characterized by the coherence peaks of the Bardeen-Cooper-Schrieffer (BCS) LDOS at $E = \pm\Delta$ (1.52 meV). The number of in-gap states observed in our experiments depends on the separation and orientation of the pairs of Gd atoms with respect to the Bi lattice, as shown in Fig. 1I, and is indicative of the nature and strength of interactions between the spins on the anisotropic Bi surface (see Supplementary Section I for detailed analysis and Sections II to IX for complete dataset). For example, a pair with separation of $2.21a$ ($a = 4.75$ Å, as one of the Bi lattice constants) shows four in-gap states, the spatial structures of which are consistent with that of bonding and anti-bonding YSR states expected for a ferromagnetically aligned pair of spins (Fig. 1J). The average energy of the two YSR states $(\Omega_1 + \Omega_2)/2$ indicates the exchange coupling strength between one of the Gd atoms and the superconducting Bi electrons, and the energy splitting between the two states $|\Omega_2 - \Omega_1|$ is induced by the wavefunction overlap between the YSR states localized at two nearby Gd atoms (30–33).

Quantum Phase Transition Controlled by Tuning Separation of Spins

Before analyzing the different possible regimes of spin alignments in our platform, we examine one sequence of experiments in which a pair of Gd atoms are brought close together systematically along the diagonal direction on the Bi lattice (black arrow in Fig. 1I). As shown in

Fig. 2 A and B, the YSR states of such a pair start at $E \sim \pm\Delta$ when the separation between spins is large. As the separation between the spins decreases, a single pair of YSR states with energies $E = \pm\Omega$ move to energies deeper inside the gap and then split into two pairs with energies $E = \pm\Omega_1, \pm\Omega_2$ at even smaller separations ($r \leq 2.77a$) with the energy splitting between the two states ($|\Omega_2 - \Omega_1|$) increasing dramatically. For $r \leq 1.38a$, the interactions between YSR states induced by the two Gd atoms are so strong that can generate a splitting larger than $\Delta$, and therefore drive one pair of the YSR states across the chemical potential at zero bias. The zero-energy crossing behavior is identified by a switch of the asymmetry in the intensity of the pair of YSR states, i.e., a switch of higher intensity state from positive to negative biases, as the two atoms move closer to each other from $r \geq 2.08a$ to $r \leq 1.38a$ (Fig. 2 A and B). The electron- and hole-like YSR states are asymmetric in intensity because they experience different local Coulomb potential from the magnetic impurity (38, 39). If we track the YSR state with higher intensity (electron-like in this case), it starts at positive energies when the separation is large ($r \geq 2.08a$) and crosses zero to negative energies when $r \leq 1.38a$ (indicated by an orange line in Fig. 2B); whereas its hole-like partner does the exact opposite (indicated by a purple line in Fig. 2B).

Theoretical works (23, 32, 33) exploring the formation of superconducting ground states when two impurity spins are interacting through a superconductor predicted that such zero-energy crossing and intensity-asymmetry switching of one pair of YSR states in LDOS measurements mark a quantum phase transition (QPT) driven by spin-spin interactions. The superconducting condensates before and after QPT are predicted to differ in parity. The transition occurs when it becomes energetically favorable to breaking a Cooper pair in the ground states so that an unpair electron can form a spin singlet with the total spin of both magnetic impurities. After QPT, the ground state parity changes from even to odd and it is accompanied by a change of the system's total spin quantum number by 1/2. A related transition occurs when the YSR states cross zero energy in a single impurity case as a function of exchange coupling strength to the superconducting host (40–42). However, the phase transition we report here is driven by the interactions between a pair of spins and controlled by their separation (23, 32, 33). In a longer chain of spins, the ability to tune the hybridization between YSR states would allow control of the YSR bands crossing the chemical potential—a key parameter for changing the band topology of the YSR states for such chains (1, 3, 6–8).

Magnetic Transition Controlled by Tuning Separation of Spins

To investigate the nature of the spin-spin interactions in our platform, we measure the interaction-induced energy splitting ($|\Omega_2 - \Omega_1|$) of YSR states for all Gd pairs as function of their displacement vector relative to the Bi surface and plot them as blue open circles in Fig. 2C (see Supplementary Sections II to VIII for complete analysis), except for the spacing of 4.75 Å where they make a spin zero pair (likely in an antiferromagnetic state) with rather weak interactions with the superconductor (Supplementary Section IX and Fig. S22). Although previous studies have shown examples of split YSR peaks indicating ferromagnetic arrangement (35–37), here we provide a detailed study of controlling spin-spin interactions as function of spacing between spins in a superconductor.

As shown in Fig. 2C, the splitting can be observed up to $r = 2.82a$. The dependence of the splitting as a function of the separation between two Gd atoms for $r \leq 2.82a$ matches that of the theoretically expected behavior (orange curve) for overlapping YSR states induced by a pair of collinearly aligned spins interacting through the superconducting host (30), from which we extract the Fermi wavelength of Bi surface states $\lambda_F = 2.53a = 12.0$ Å. More specifically, for separations $r < \lambda_F$ and $\lambda_F < r \leq 2.82a$, the splitting behavior is expected for two Gd spins in ferromagnetic (FM) ground state; whereas for $r = \lambda_F$, the splitting always vanishes, which is consistent with two Gd spins in antiferromagnetic (AFM) ground state and their YSR states wavefunctions are orthogonal to each other. The oscillatory behavior indicates the spin-spin interactions for $r \leq 2.82a$ are dominated by RKKY interaction (23).

For $r > 2.82a$, the significant deviation from RKKY behavior shown in Fig. 2C implies other interactions needs to be taken into account in this range. For two Gd atoms with separation $r > 2.82a$, or more specifically for an isolated Gd atom, we only observe YSR states at energies very close to the gap edge despite of Gd's large spin (for example, Fig. 2A, $r = 4.84a$ and Fig. S2E). We distinguish the weakly bound YSR states induced by an isolated Gd atom from the BCS coherence peaks using a superconducting tip that allows higher-resolution measurements (Supplementary Section X and Fig. S23) or by the asymmetric intensity of electron- and hole-like partner states measured with a normal tip (Supplementary Section I and Fig. S3E). The low binding energy of this YSR state is a clear indication of small exchange interaction ($J\mathbf{S}\cdot\mathbf{\sigma} \ll \Delta$, where $J$ is the exchange coupling constant) between Gd spin ($\mathbf{S}$) and Bi surface electrons ($\mathbf{\sigma}$) that are in-plane spin polarized due to the large Rashba type spin-orbit coupling (21). The small exchange interaction ($J\mathbf{S}\cdot\mathbf{\sigma} \sim 0$) is consistent with the out-of-plane spin polarization of the isolated Gd atom due to strong surface magnetic anisotropy, which is common for large spin magnetic atoms on an atomically ordered surface, where the full rotational symmetry is broken (43, 44).

The overall behavior of YSR states hybridization as a function of the separation between two spins can be understood, if we consider the interplay between RKKY interaction, the surface magnetic anisotropy, and the spin-orbit coupling in the substrate. For a pair of Gd spins within separation range $2.82a < r \ll \xi$, they may align in FM or AFM arrangement pointing perpendicular to the surface, therefore only weak YSR states near gap edge are observed, due to weak exchange interaction between Gd spin and in-plane spin-polarized Bi electrons. At closer separations $r \leq 2.82a$, we expect the distance-dependent RKKY interaction overcomes the surface magnetic anisotropy and makes the Gd pair spins lie in-plane to lower the total energy, leading to strong exchange interaction with the Bi surface electrons that drives the YSR states deep inside the gap. Depending on the sign of the RKKY interaction, both in-plane FM alignment and in-plane AFM alignment of the spins can be the ground state of the system. Here we find in-plane FM states with four YSR states are at $r < \lambda_F$ (Fig. 1 E and F). In addition, close to the alignment transition at $r \sim \lambda_F$, we also find evidence for in-plane AFM states with two degenerate (zero-splitting) YSR states (for example, at $E = \pm\Omega$ in Fig. 1H), with intensities equal to the sum of split YSR states intensities of FM aligned spins at slightly closer spacing (at $E = \pm\Omega_1, \pm\Omega_2$ in Fig. 1F).

In the crossover region with separation $r \sim \lambda_F$, four particular pair configurations ($r = 2.16a$, $2.21a$, $2.77a$, and $2.82a$) show variability in their spin arrangements from location to location on the surface, sometimes displaying splitting of YSR states consistent with FM alignment, or no splitting at all as expected for AFM alignment (Supplementary Section VI, Figs. S13 and S14). More importantly, for distances larger than $\lambda_F$, the interplay between RKKY interaction, spin-orbit coupling, and surface magnetic anisotropy makes the spin alignment of the pairs more complex.

Theoretical Model

A theoretical model that captures the interplay between the surface anisotropy and the RKKY interaction in the presence of strong spin-orbit coupling can be used to qualitatively describe the changing spin alignment in our system. Since we are considering interactions between spins on length scale far below the superconducting coherence length ($r \ll \xi$), the influence of superconductivity on spin-spin interactions will not be significant. Furthermore, recent work (22, 34) considering the contribution of the overlap of the YSR state to spin-spin interactions also shows that RKKY interaction is dominant on length scale below $r < \sqrt{\xi \lambda_F / \pi}$, which is 53 Å for our system. Therefore, for simplicity we consider a pair of spins interacting via RKKY interaction through a two-dimensional electron gas with strong spin-orbit interaction in the presence of magnetic surface anisotropy (see Supplementary Section XI for details). For large separations of the spins, with sufficiently large surface anisotropy term $H_A = -AS_z^2$, where $A > 0$ and $S_z$ is the spin component perpendicular to the surface, the ground state of this system is clearly either non-interacting or weakly FM or AFM with spins pointing perpendicular to the surface. As the separation between the spins becomes of the order of $\lambda_F$, the RKKY interaction begins to dominate the properties of this system, resulting in a small region of in-plane AFM aligned pairs and finally at $r < \lambda_F$, in-plane FM aligned pairs with their spins pointing along the direction perpendicular to the line connecting them (Fig. 2D). Transition between out-of-plane to in-plane spin alignments in pairs underlies our observation that YSR states of such pairs are far lower in energy than a single Gd atom. The transition is also consistent with the presence of degenerate YSR states with no splitting for Gd pairs with $r \sim 2.5a$, because in-plane Rashba type spin-orbit coupling can lift the degeneracy and lead to YSR states splitting for a pair of spins with out-of-plane AFM alignment, as shown in a recent experiment (45). However, if the pair of spins are in the in-plane AFM ground state and aligned perpendicularly to the connection between them (Fig. 2D, $r \sim 2.5a$), the degeneracy can still be preserved (see Supplementary Section XII for details).

Experiments on Larger Spin Assemblies

To demonstrate that changing of spin alignment in larger assemblies opens up the possibility of engineering the band structure of YSR states in our platform, we constructed structures made of three and four Gd atoms. Figure 3 shows that adding a third atom to a pair of in-plane FM aligned spins with separation $r < \lambda_F$ (Fig. 3 A–C) splits their overlapping YSR states further to create a 6-peak structure with varying intensity on the three different atoms (Fig. 3 E–H). A minimal model (33) of the overlap between the YSR states ($\Omega_0$) with nearest ($t_1$) and next-nearest neighbors' matrix elements ($t_2$), as shown in Fig. 3D can be used to understand the 6-peak structure for such 3-atom spin chains (Fig. 3 E–L). The comparison of the energy splitting of the YSR peaks to the model calculation, however, reveals that for a 3-atom spin chain with $2a$ spacing, the ratio $t_2/t_1$ is surprisingly smaller than what expected for three atoms in FM alignment. Figure 2C (orange curve) shows that YSR overlap for $2a$ and $4a$ distances should yield $t_2 \sim -0.8 t_1$ for FM alignment, while to capture the data in Fig. 3 we have to use 5 to 10 times smaller $t_2/t_1$ ratio (Supplementary Section XIII). This indicates that the three spins are not collinear, which would reduce the overlap interactions by a factor of $\cos(\theta/2)$, where $\theta$ is the angle between the nearest neighbor spins. From this simple model, we suggest that spins in a 3-atom chain prefer to make an in-plane non-collinear alignment (Supplementary Section XIII). Theories predict helical spin order to be a natural consequence of RKKY interaction for an infinite chain in one dimension (2), and could be stabilized in the presence of large spin-orbit coupling in two dimensions (46). Our experimental data show that the magnetic frustration can lead to non-collinear spin ground states in a 3-atom chain, and potentially helical order when longer chains are constructed. Theoretical studies have established that when the wavelength of helical order is large than $\lambda_F$, then such chains will be in a topological phase hosting MZMs at their ends (1–6, 8).

Examining the properties of a 3-atom spin chain when a fourth atom is brought into its close proximity shows that spin-spin interactions can be mediated and controlled in the superconducting state on length scales much larger than the lattice spacing or $\lambda_F$ in our system. As shown in Fig. 4, the 6-peak structure of the 3-atom chain responds to the presence of the fourth atom as it couples to the chain starting from ~ 16 Å away from the chain. In effect, a spin (labeled as Atom 1 in Fig. 4B) is getting influenced by another spin (Atom 4 in Fig. 4B) via the chain even if they are more than 26 Å apart. Compared to previous experiments where spins in a chain are coupled in a collinear manner through exchange interaction (47, 48), here the combination of long-range spin-spin exchange coupling through chains and the possibility to modulate the YSR band structure through tuning the YSR states hybridization make it possible to quantitatively control the topological phase transition of complex spin structures on a superconducting surface. With the tunability of our platform, it is possible to create well-developed helical spin states in a topological phase in longer chains which host MZMs at their ends. Detailed studies of MZMs coupling to each other as a function of chain length, as well as coupling MZMs to nearby spins can be performed in this system. Finally, it would be possible to create complex coupled chains in which $Z_2$ symmetry or fusion of MZMs can be directly tested in our platform.

**Materials and Methods**

Both sample preparation and *in situ* scanning tunneling microscopy (STM) measurements were performed in a home-built dilution refrigerator STM (base temperature: 250 mK) equipped with ultra-high vacuum (UHV) preparation chambers (base pressure: $2.0 \times 10^{-10}$ Torr) (18).

The Nb(110) single crystal was prepared by argon ion sputtering and subsequent flash annealing (10 cycles, 30 seconds each) at 1600 °C to form clean and flat surface. Nominally 5 monolayer (ML) Bi (99.9999%) films were evaporated from standard Knudsen cells onto Nb(110) surface which was held at –150 °C by liquid nitrogen cooling. Fully covered 4 ML Bi films and partially covered 6 ML Bi films (Fig. S1 A and B) were formed uniformly after being annealed at 100 °C for 30 minutes and exhibit a hard superconducting gap ($\Delta = 1.52$ meV) (Fig. S1C) which is nearly identical to that of Nb due to the proximity effect (proximity effect becomes weaker once Bi films are thicker than 6 ML). Sub-monolayer Gd (99.9%) single atoms were then deposited on Bi surface at cryogenic temperature (~ 40 K) (Fig. S1D).

All STM measurements were performed on 6 ML Bi films at effective electron temperature of 250 mK (calibrated by measurements of the superconducting gap, Fig. S1C) using a sharp W tip. The tip was treated by field emission and controlled indention on Cu(100) surface until it was atomically sharp. Topographic images were taken using constant current mode with closed feedback loop at sample bias voltage $V_s = -1$ V and tunneling current $I_t = 10$ pA. Differential conductance (*dI/dV*) spectra were acquired at either $V_s = -4$ mV, $I_t = 40$ pA or $V_s = -20$ mV, $I_t = 200$ pA under open feedback condition, using a lock-in amplifier at a frequency of 712.9 Hz with AC rms excitation $V_{rms} = 20$ μV.

Gd atoms were manipulated by the tip laterally with closed feedback loop at $V_s = \pm 1.5$ mV, $I_t = 200$ pA. An example of building Gd pairs using atomic manipulation is shown in Fig. S2.

**Data Availability:** The data that support the findings of this study are available within the paper and the Supplementary Information.

**Acknowledgments:** We acknowledge discussions with Y. Meir. This work has been primarily supported by Gordon and Betty Moore Foundation as part of EPiQS initiative (GBMF4530 and GBMF9469), ONR-N00014-17-1-2784, ONR-N00014-14-1-0330, NSF-MRSEC programs through the Princeton Center for Complex Materials DMR-142054 and NSF-DMR-2011750 grants, as well as the NSF-DMR-1608848, NSF-DMR-1904442 grants. Also supported by the Swiss National Science Foundation and NCCR QSIT (J.K., O.D., S.H., and D.L.) and the European Union's Horizon 2020 research and innovation program (ERC Starting Grant, grant agreement No 757725) (J.K.). A.Y. acknowledges the hospitality of the Trinity College and Cavendish Laboratory in Cambridge, UK during the preparation of this manuscript, which was also funded in part by a QuantEmX grant from the Institute for Complex Adaptive Matter and the Gordon and Betty Moore Foundation (GBMF5305).

**Author Contributions:** H.D., Y.H., M.T.R., and A.Y. designed and conducted the experiment and data analysis; S.H., O.D., J.K., and D.L. designed and performed the theoretical calculations; all authors contributed to the writing of the manuscript.

**Competing Interest Statement:** The authors declare no competing interests.

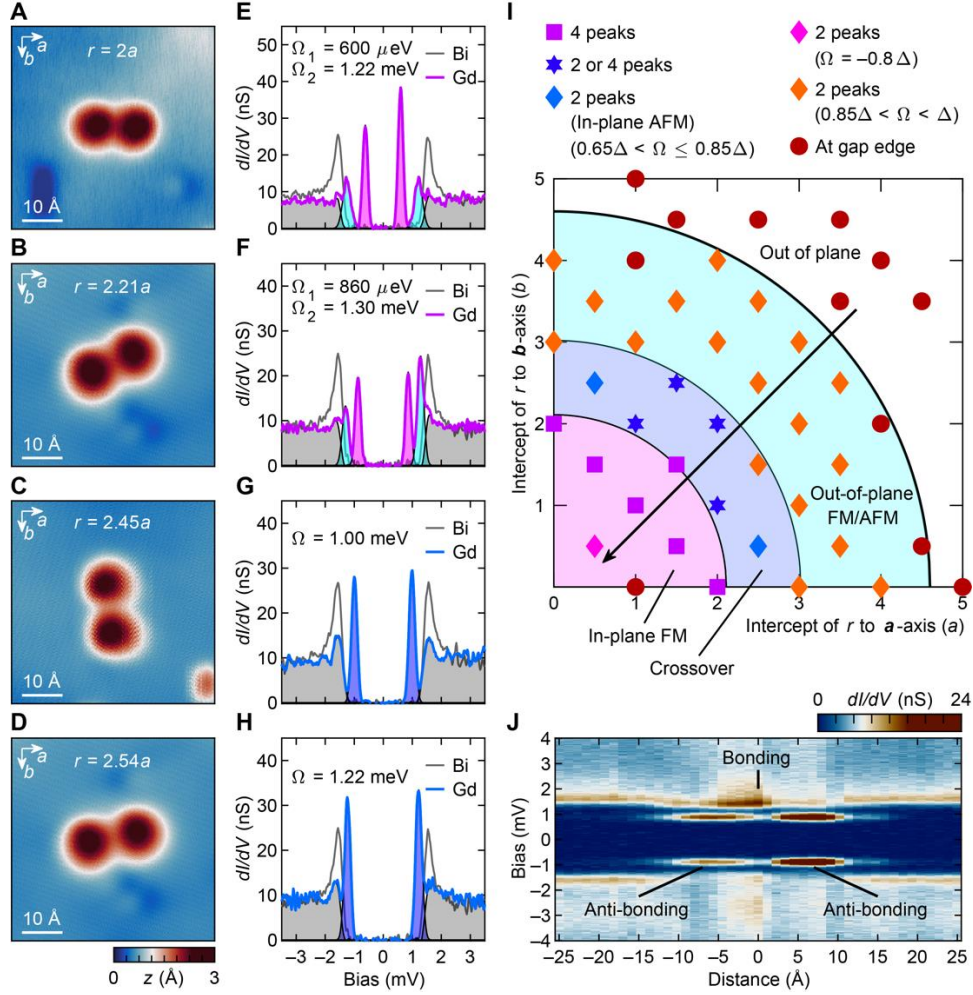

**Fig. 1.** YSR states and magnetic ground states of Gd pairs on proximitized superconducting Bi surface. (**A**–**D**) Four examples of Gd pairs with different interatomic distances and orientations with respect to Bi lattice. *a* and *b* are the crystallographic directions of the Bi surface with lattice constants $a = 4.75$ Å and $b = 4.55$ Å (see Fig. S1B). (**E**–**H**) $dI/dV$ spectra taken on the Gd pairs in panels A–D, respectively. The gray curves in the background are spectra taken on Bi surface away from Gd atoms, showing the superconducting gap ($\Delta = 1.52$ meV) for reference. The YSR states energies $\Omega$ (or $\Omega_1$, $\Omega_2$) are measured from the electron-like in-gap states with higher spectral intensity. The spectral weights of bonding and anti-bonding YSR states are color coded by cyan and magenta; whereas that of the degenerate YSR states are color coded by blue. After

subtracting all YSR states from the whole spectra, the rest of spectral weights are color coded by gray, showing the suppressed coherence peaks (see Supplementary Section I for detailed analysis). (**I**) A phase diagram of the magnetic ground states of Gd pairs revealed by the YSR states analysis (see Supplementary Sections II to IX for complete analysis). (**J**) A linecut of *dI/dV* spectra taken across a Gd pair with $r = 2.21a$ showing the spatial distribution of bonding and anti-bonding YSR states.

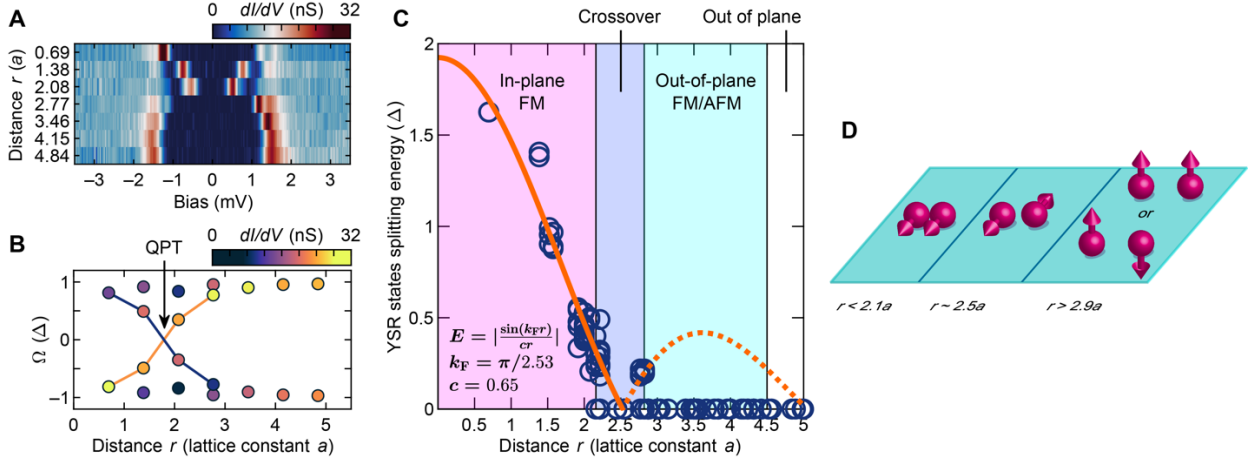

**Fig. 2.** Quantum phase transition and magnetic transition controlled by tuning the distance between two Gd atoms. (**A**) *dI/dV* spectra of Gd pairs with configurations along the black arrow indicated in Fig. 1I. (**B**) A color plot of the YSR states intensity extracted from panel A. The orange (purple) line indicates evolution of electron (hole)-like YSR states. The crossing of orange and purple lines implies a quantum phase transition (QPT). The QPT occurs at $r \sim 1.55a$ (for example, Figs. S6C, S7D, S22 A and D), closer separations comparing to the pair in Fig. 1A. (**C**) The energy splitting between bonding and anti-bonding YSR states $|\Omega_2 - \Omega_1|$ (blue open circles) as a function of interatomic distance $r$. The splitting is zero if there is only one pair of YSR states. See Supplementary Sections II to VIII for complete analysis. The orange curve is a fit to $E = |\frac{\sin(k_F r)}{cr}|$, which is the theoretically predicted YSR states splitting for a pair of collinearly aligned spins as a function of their separation. Here $c = \frac{k_F(1+\alpha^2)}{4\alpha\Delta}$, where $\alpha = \pi S J \nu_F$ characterizes the single magnetic atom induced YSR state; $J$ is the exchange coupling between the spin and superconducting electrons; $\nu_F$ is the density of states at Fermi level in the normal state of the superconductor. (**D**) A schematic showing the magnetic ground state of Gd pairs can be tuned from out-of-plane FM/AFM to in-plane AFM and then to in-plane FM phase as moving two Gd atoms closer.

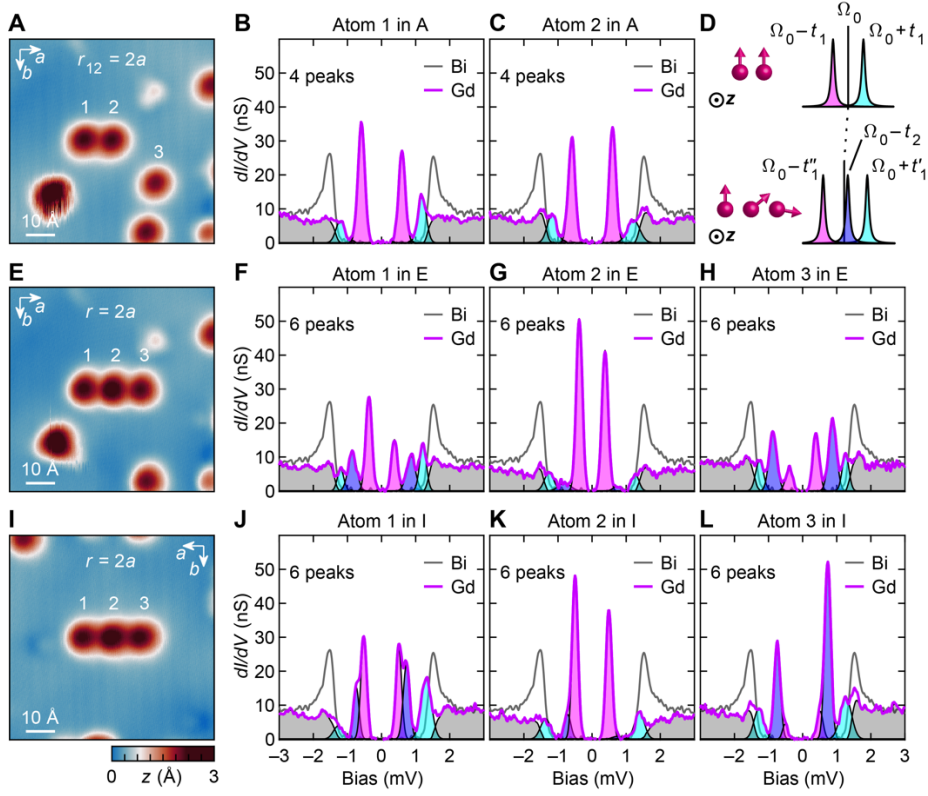

**Fig. 3.** YSR states and non-collinear magnetic ground states of 3-Gd-atom chains. (**A–C**) Topograph and *dI/dV* spectra of a Gd pair with $r = 2a$, showing it is in in-plane FM state with two pairs of YSR states. (**D**) Extracting YSR states coupling parameters ($\Omega_0$, $t_1$, and $t_2$) from Gd pair in panel A and 3-atom chain in panel E. $\Omega_0$ is the average energy of all electron-like YSR states; $t_1$ is the nearest neighbor coupling matrix element; $t_2$ is the next-nearest neighbor coupling matrix element between Atom 1 and Atom 3 in panel E. $t'_1 = (\sqrt{8t_1^2 + t_2^2} + t_2)/2$, $t''_1 = (\sqrt{8t_1^2 + t_2^2} - t_2)/2$, for clarity. The small value of $t_2$ implies that the coupling between Atom 1 and Atom 3 in panel E is reduced by the non-collinear alignment of spins. (**E–H**) Topograph and *dI/dV* spectra of a 3-atom chain evenly spaced with $r = 2a$, built by moving Atom 3 to the right side of Atom 2 in panel A. The magenta, blue and cyan coded peaks are bonding (0 node), non-bonding (1 node), and anti-bonding (2 nodes) YSR states from further splitting after adding Atom 3. (**I–L**) Topograph and *dI/dV* spectra of another example of 3-atom chain with $r = 2a$,

built at a different location showing slightly different YSR states energies but similar behavior as the one in panel E.

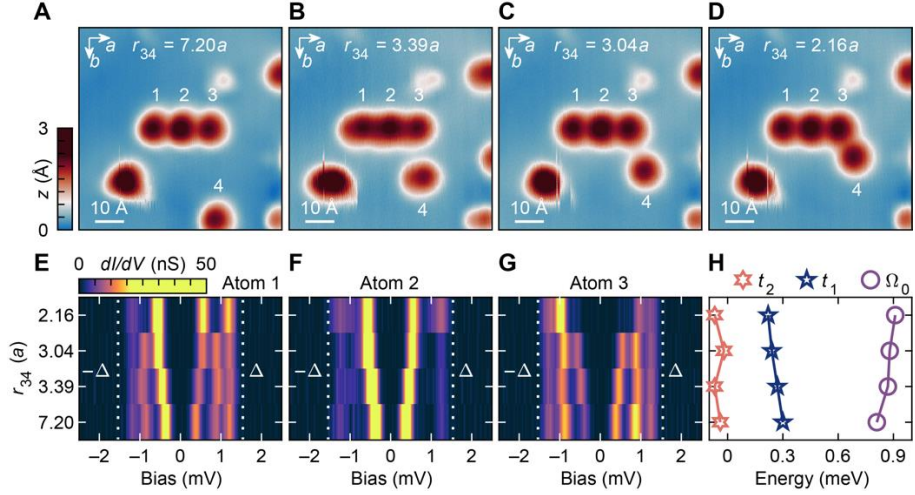

**Fig. 4.** Long-range spin-spin interactions between a 3-Gd-atom chain and a Gd atom. (**A–D**) A sequence of atomic manipulation moving Atom 4 from the lower right corner towards Atom 3 step by step. Atoms 1, 2 and 3 are evenly spaced with $r = 2a$. $r_{34}$ is the distance between Atom 3 and Atom 4. (**E–G**) Color plots of $dI/dV$ spectra taken on the Atoms 1 to 3 in panels A–D, from bottom to top, showing YSR states evolution through the manipulation sequence. The density of states outside of the gap ($|E| \geq \Delta$) are subtracted for clarity (see Figs. S26 and S27 for raw data). (**H**) The YSR coupling parameters ($\Omega_0$, $t_1$, and $t_2$) extracted from panels E–G (see Supplementary Section XIV). Their evolution implies the spin alignment of the 3-atom chain (Atoms 1 to 3) is changing through the manipulation sequence, due to the perturbation of Atom 4 via long-range spin-spin interactions.

# Supplementary Information for

# Tuning interactions between spins in a superconductor

Hao Ding[1†], Yuwen Hu[1†], Mallika T. Randeria[1‡], Silas Hoffman[2,3], Oindrila Deb[2], Jelena Klinovaja[2], Daniel Loss[2], Ali Yazdani[1]*

[1]Joseph Henry Laboratories & Department of Physics, Princeton University, Princeton, NJ 08544, USA.
[2]Department of Physics, University of Basel, Klingelbergstrasse 82, CH-4056 Basel, Switzerland.
[3]Department of Physics, University of Florida, Gainesville, FL 32611, USA.

*To whom correspondence may be addressed. Email: yazdani@princeton.edu
[†]These authors contributed equally to this work.
[‡]Present address: Department of Physics, Massachusetts Institute of Technology, Cambridge, Massachusetts 02139, USA.

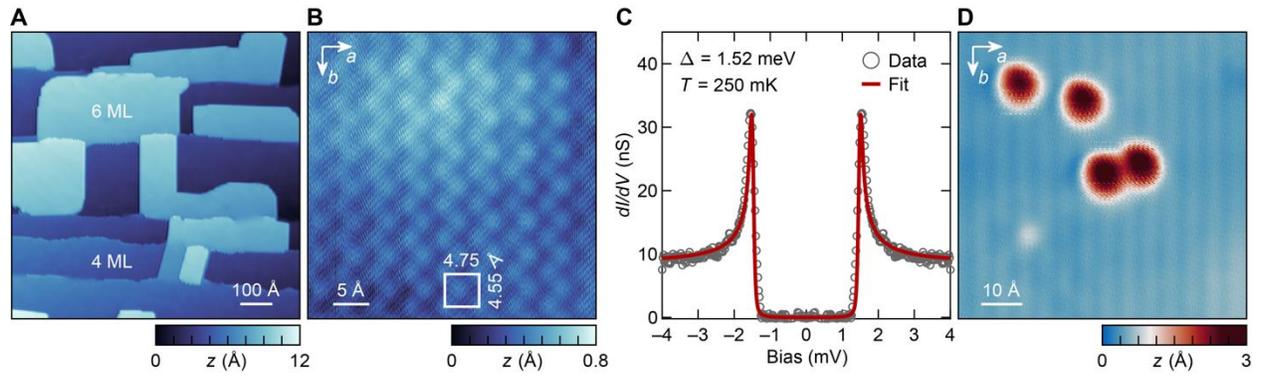

**Fig. S1.** Characterization of Gd atoms on Bi/Nb(110) heterostructure. (**A**) A large-scale topography of both fully covered 4 monolayers (ML) and partially covered 6 ML Bi films epitaxially grown on Nb(110). (**B**) A zoom-in topography taken on a 6 ML Bi film showing the typical zig-zag atomic structures of Bi surface with lattice constants $a = 4.75$ Å and $b = 4.55$ Å. (**C**) $dI/dV$ spectrum (gray circles) taken on a 6 ML Bi film exhibiting a hard superconducting gap and BCS fit (red curve). (**D**) Topography of the Bi surface after sub-monolayer of Gd atoms deposition showing two individual Gd atoms and a Gd pair sitting on Bi surface.

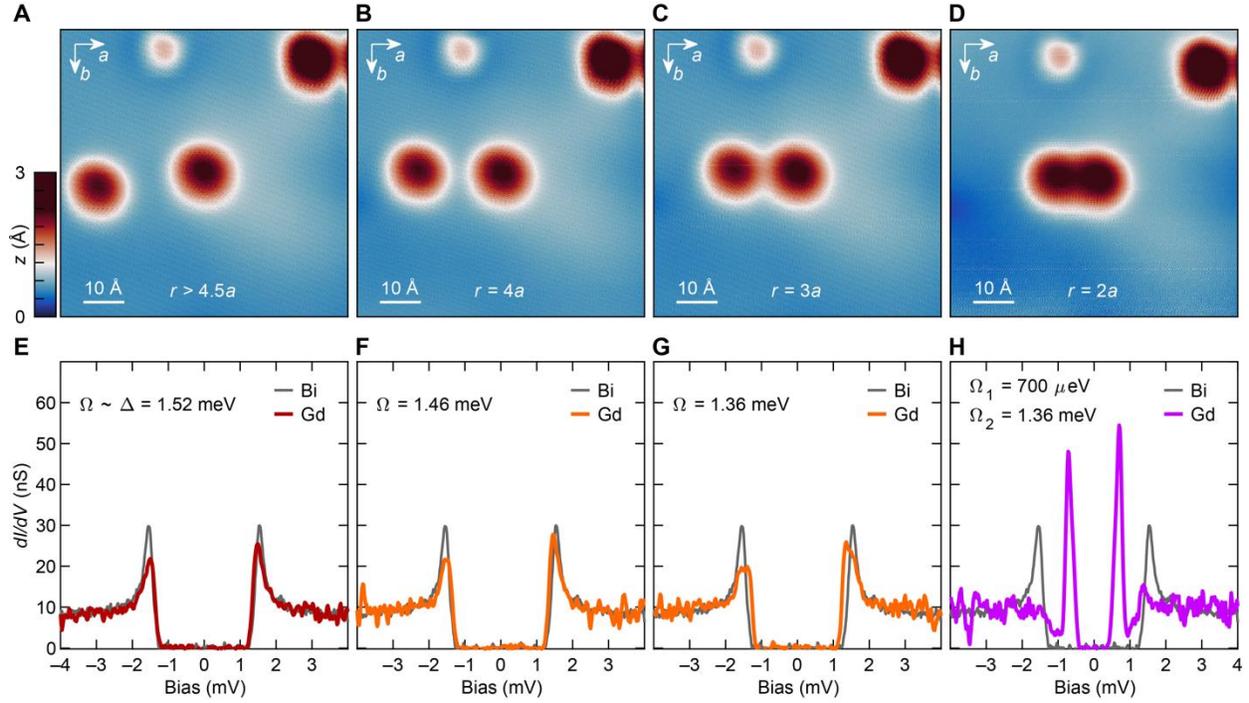

**Fig. S2.** An example of Gd pair built by atomic manipulation. (**A**–**D**) A sequence of atomic manipulation moving a single Gd atom from the left towards another Gd atom at the center step by step, with their separation ($r$) decreasing from $r > 4.5a$ to $r = 4a$, $3a$, and $2a$, respectively. $a$ and $b$ are the crystallographic directions of the Bi surface with lattice constants $a = 4.75$ Å and $b = 4.55$ Å, respectively (see Fig. S1B). (**E**–**H**) $dI/dV$ spectra taken on the Gd atom at center in panels A–D, respectively. The YSR states energies ($\Omega$) are measured from the electron-like in-gap states with higher spectral intensity (see Supplementary Section I). The gray curves in the background are spectra taken on 6 ML Bi surface away from Gd atoms, showing the superconducting gap ($\Delta = 1.52$ meV) for reference.

**Supplementary Text**

Section I. YSR states analysis

As shown in Figs. 1 and S2, the YSR states induced by the Gd pairs vary in energy dramatically depending on both distances between the two atoms and their orientations with respect to the Bi lattice underneath. We find almost all the YSR states can be categorized into 5 cases (Fig. S3), and with a 3-step analysis, we can extract the YSR states energies, therefore obtain information about the spin ground states of Gd pairs. We describe this procedure and the color codes used to describe its outcome in the various figures below.

**Step 1.** Compare the *dI/dV* spectra taken on Gd pairs (red curves) to spectra measured on Bi surface away from Gd atoms (gray curves) in Fig. S3 A–E. Although only the first three (Fig. S3 A–C) show clear in-gap states, all five of them show particle-hole asymmetry signatures (indicated by black arrows). The particle-hole asymmetry is a hallmark of presence of YSR states. Unlike the coherence peaks of the superconducting gap which are particle-hole symmetric, the particle-hole symmetry of YSR states is broken due to the interaction between the magnetic impurity and electron (hole)-like quasi-particles (1, 2). In the case of single Gd atoms on Bi/Nb(110), the electron-like quasi-particles have higher spectral intensity than the hole-like quasi-particles. Therefore, even though the last two spectra (Fig. S3 D and E) do not show obvious in-gap states, we can still identify the presence of YSR states at energies near the gap edge.

**Step 2.** For spectra that show YSR states near the gap edge (Fig. S3 D and E), extract the energies of electron-like YSR states ($\Omega$) with higher spectral intensities (hole-like YSR states energies are therefore at $-\Omega$), and color code the spectra orange if $0.85\Delta < \Omega < \Delta$ (Fig. S3I), or red if $\Omega \sim \Delta$ (Fig. S3J). For spectra that show YSR states which are well separated in energy from the gap edge (Fig. S3 F–H), fit the two peaks deep in the gap with Gaussian function and color code the spectral weights of the peaks with magenta. Extract the energies of electron-like YSR states with higher spectral weights at Ω (hole-like YSR states energies are therefore at –Ω). Then subtract the two peaks from the spectra and color code the rest of spectral weights with gray. Examine if the gray coded spectra are particle-hole symmetric. If they are symmetric, there is no more YSR state (Fig. S3G); otherwise, there are more near the gap edge (indicated by black arrows in Fig. S3 F and H).

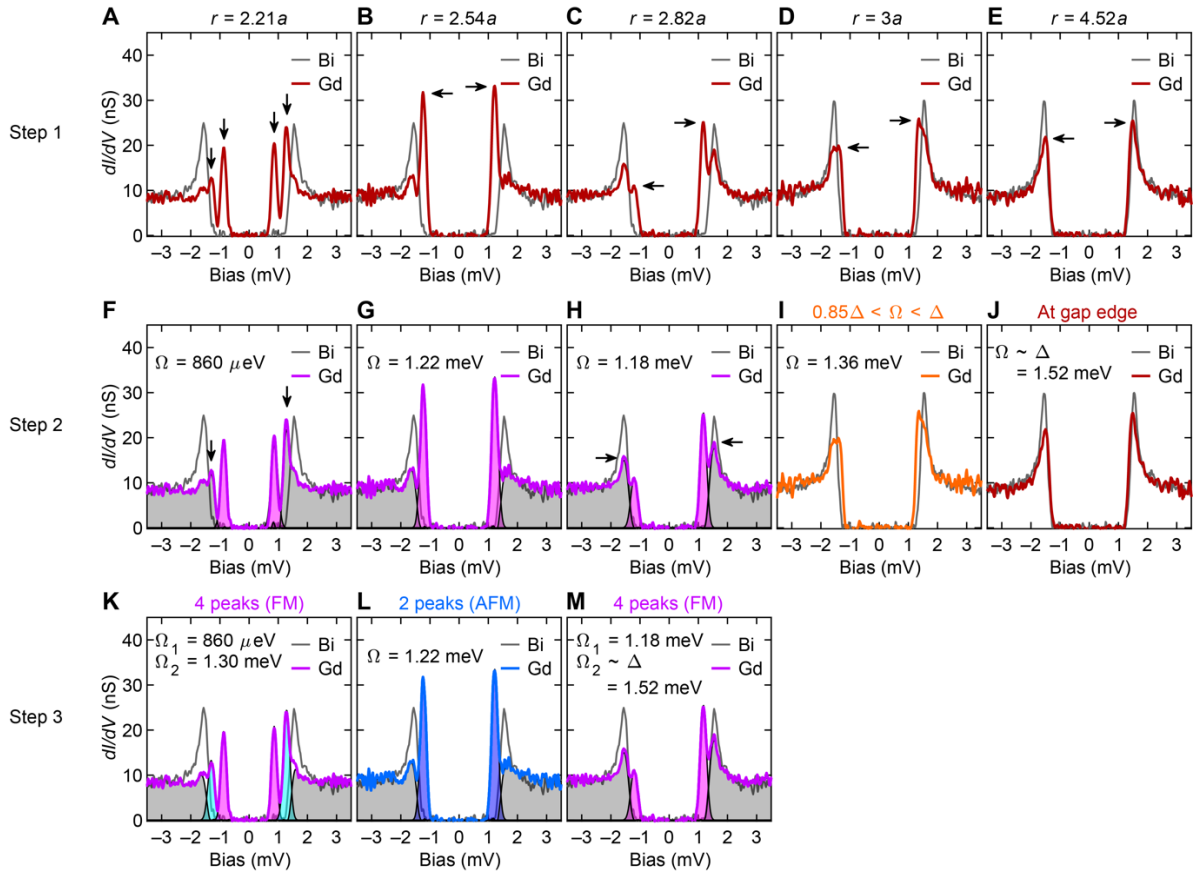

**Fig. S3.** YSR states analysis procedure for 5 typical Gd pairs with different configurations. $r$ is the separation between two Gd atoms. $a = 4.75$ Å, as one of the Bi lattice constants.

**Step 3.** In the case of gray coded spectral weights being asymmetric and well separated from the gap edge (Fig. S3F), fit the other two peaks and color code the spectral weights of the peaks with cyan; extract the energy of electron-like YSR states ($\Omega_2$) with higher spectral weights; then subtract the four peaks from the spectra and color code the rest of spectral weights with gray (Fig. S3K). Now the gray coded spectral weights are symmetric, therefore there is no more YSR state. In the case of gray coded spectral weights being asymmetric but inseparable from the gap edge (Fig. S3H), extract the energy of electron-like YSR states ($\Omega_2$) with higher spectral weights. So, we label both cases (Fig. S3 K and M) as in FM states with two pairs of YSR states corresponding to the bonding and anti-bonding states. In the case of gray coded spectral weights being symmetric (Fig. S3G), one can find that the spectral weights of the YSR peaks are nearly twice as high as peaks in Fig. S3F, which suggests it is in an AFM state with one pair of doubly degenerate YSR states. To distinguish it from others, we change its color code to blue (Fig. S3L).

Section II. Ferromagnetic YSR states with splitting $|\Omega_2 - \Omega_1| > \Delta$

For Gd pairs with close atomic separations $r = 0.69a$ (Fig. S4) and $1.38a$ (Fig. S5), one of the electron-like YSR states ($\Omega_1$) with highest spectral intensity always resides at a negative energy. It indicates that the wave function hybridization induced energy splitting between bonding and anti-bonding YSR states $|\Omega_2 - \Omega_1|$ is larger than $\Delta$, and therefore pushes $\Omega_1$ across zero energy to the negative side. Compared to Gd pairs with both $\Omega_1$ and $\Omega_2$ at positive energies (for example, Fig. S3K), the superconducting ground state is predicted to be different by the presence of an unpaired spin, so the zero crossing behavior of the YSR states marks a quantum phase transition (3–5).

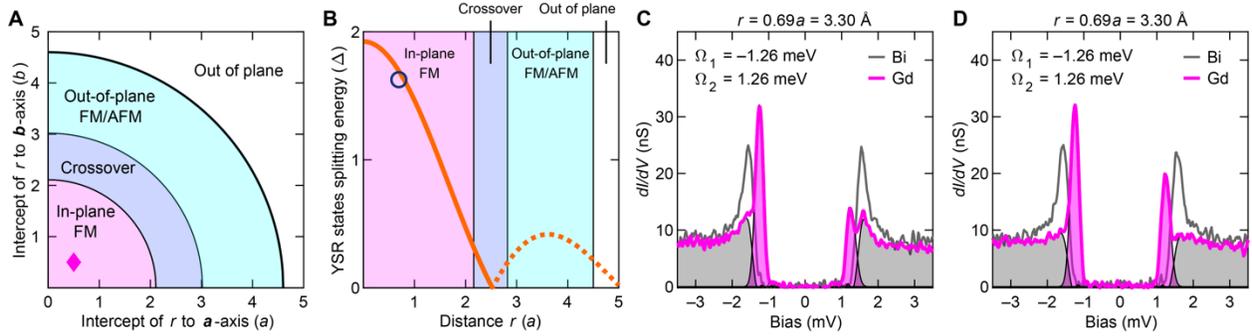

**Fig. S4.** YSR states of Gd pairs with separation $r = 0.69a$. (**A**) Its configuration in the phase diagram as in Fig. 1I. (**B**) YSR states splitting energies in the phase diagram as in Fig. 2C. (**C** and **D**) $dI/dV$ spectra taken on Gd pairs with separation $r = 0.69a$ at different locations on Bi films. $a = 4.75$ Å, as one of the Bi lattice constants.

In the case of $r = 0.69a$ (Fig. S4), although there are only one pair of degenerate peaks inside the gap, the particle-hole asymmetry of the peaks is opposite to the AFM case in Fig. S3L. This behavior can only be explained by $\Omega_1 = -\Omega_2$, i.e., one of the electron-like states ($\Omega_1$) with higher spectral intensity is pushed below zero energy and meets the hole-like states corresponding to the other electron-like states ($\Omega_2$) at energy $-\Omega_2$. So, the doubly degenerate peaks for $r = 0.69a$ (Fig. S4 C and D) are from degeneracy of one electron-like state and one hole-like state, whereas in the AFM case (Fig. S3L), the degeneracy is from two electron (hole)-like states.

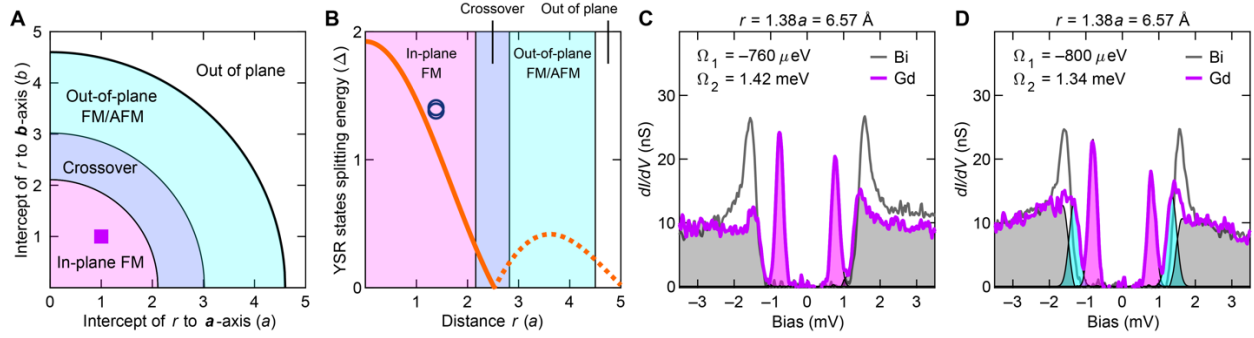

**Fig. S5.** YSR states of Gd pairs with separation $r = 1.38a$. (**A**) Its configuration in the phase diagram as in Fig. 1I. (**B**) YSR states splitting energies in the phase diagram as in Fig. 2C. (**C** and **D**) $dI/dV$ spectra taken on Gd pairs with separation $r = 1.38a$ at different locations on Bi films. $a = 4.75$ Å, as one of the Bi lattice constants.

Section III. Ferromagnetic YSR states with splitting $|\Omega_2 - \Omega_1| \sim \Delta$

For Gd pairs with atomic separations $r = 1.52a$ (Fig. S6C) and $1.57a$ (Fig. S7D), both electron-like and hole-like states of one YSR state ($\Omega_2$) meet at the zero energy, which marks the quantum critical point (QCP) of the quantum phase transition (QPT). This is driven by the splitting $|\Omega_2 - \Omega_1|$ between bonding and anti-bonding YSR states. It happens when splitting $|\Omega_2 - \Omega_1|$ shifts $|\Omega_1|$ or $|\Omega_2|$ to zero energy.

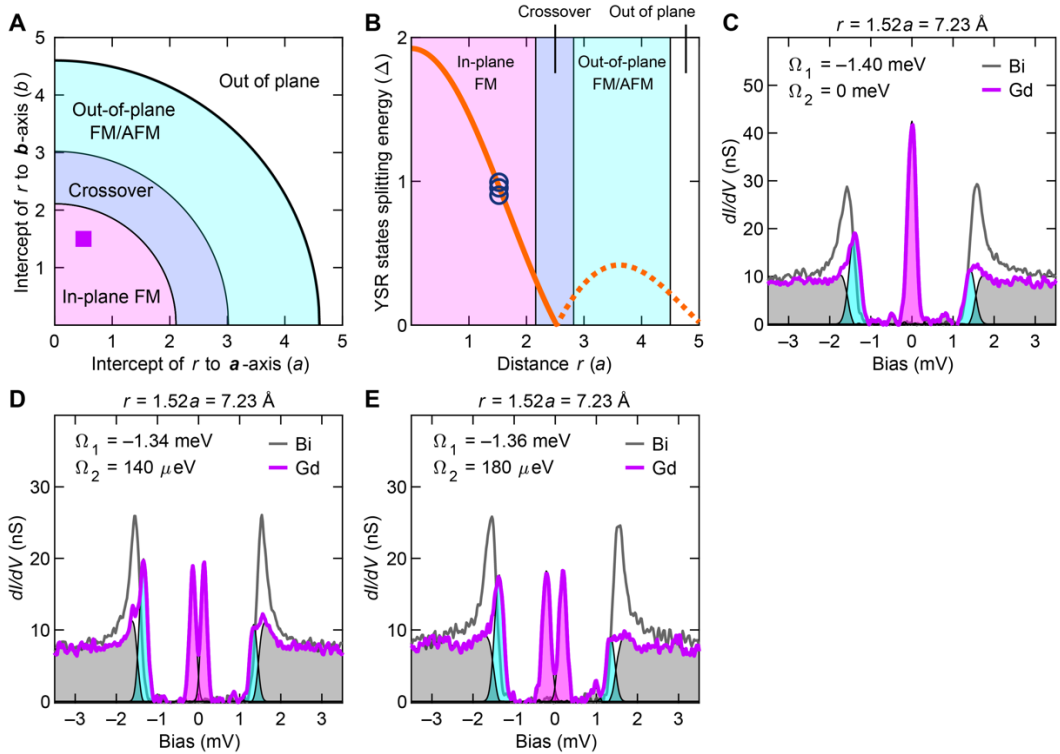

**Fig. S6.** YSR states of Gd pairs with separation $r = 1.52a$. (**A**) Its configuration in the phase diagram as in Fig. 1I. (**B**) YSR states splitting energies in the phase diagram as in Fig. 2C. (**C**–**E**) $dI/dV$ spectra taken on Gd pairs with separation $r = 1.52a$ at different locations on Bi films. $a = 4.75$ Å, as one of the Bi lattice constants.

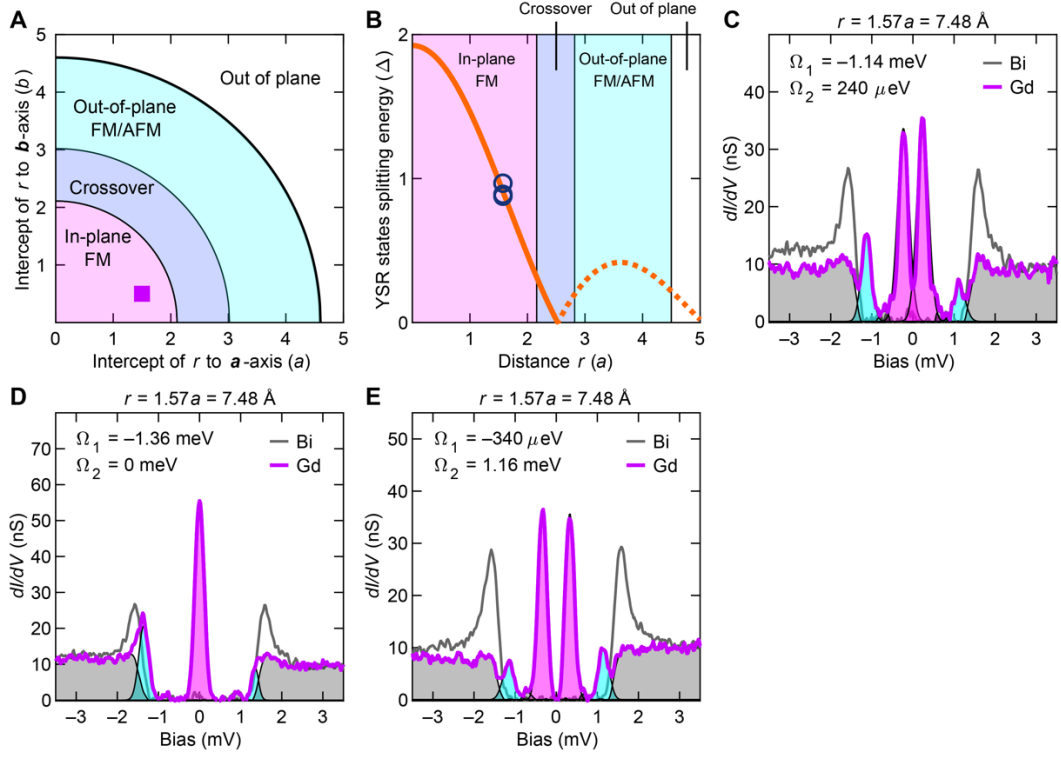

**Fig. S7.** YSR states of Gd pairs with separation $r = 1.57a$. (**A**) Its configuration in the phase diagram as in Fig. 1I. (**B**) YSR states splitting energies in the phase diagram as in Fig. 2C. (**C**–**E**) $dI/dV$ spectra taken on Gd pairs with separation $r = 1.57a$ at different locations on Bi films. $a = 4.75$ Å, as one of the Bi lattice constants.

Section IV. Ferromagnetic YSR states with splitting $|\Omega_2 - \Omega_1| < \Delta$

For Gd pairs with atomic separations $r = 1.92a$ (Fig. S8), $2a$ (Fig. S9), and $2.08a$ (Fig. S10), one of the YSR states is very deep in the gap (~ $0.4\Delta = 600$ μeV), which makes it more localized and exhibit much higher spectral intensity at most of the time. This also makes it more difficult to distinguish electron-like states and hole-like states since their intensities are almost equally strong.

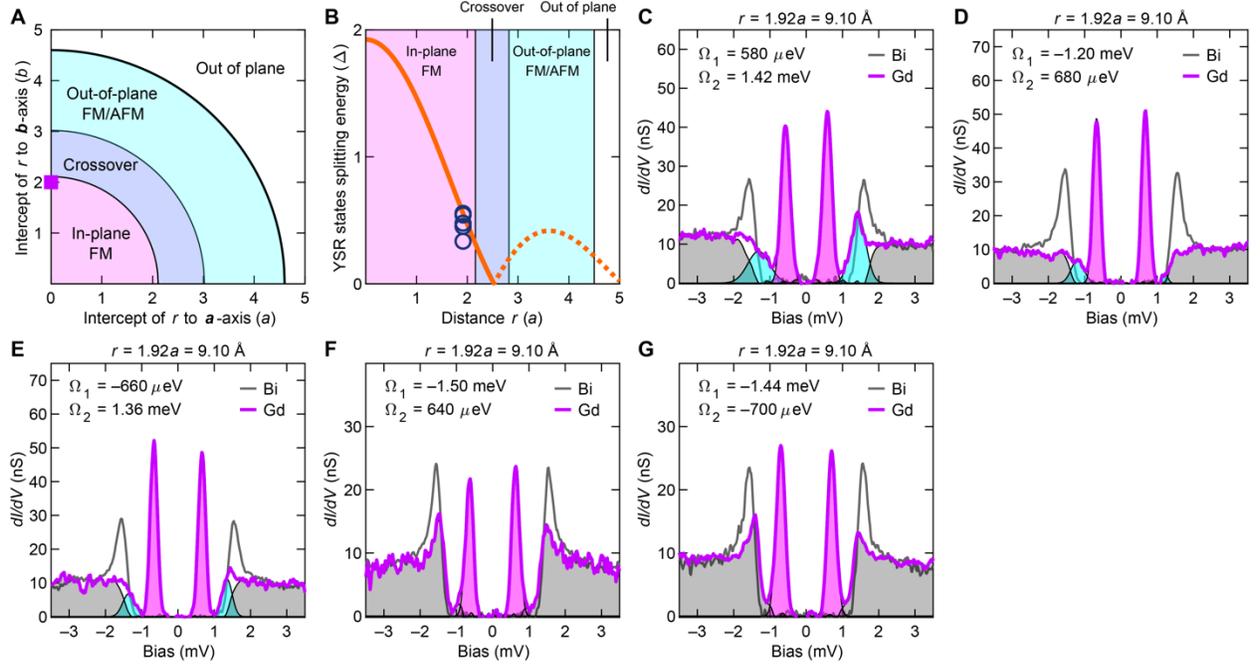

**Fig. S8.** YSR states of Gd pairs with separation $r = 1.92a$. (**A**) Its configuration in the phase diagram as in Fig. 1I. (**B**) YSR states splitting energies in the phase diagram as in Fig. 2C. (**C**–**G**) $dI/dV$ spectra taken on Gd pairs with separation $r = 1.92a$ at different locations on Bi films. $a = 4.75$ Å, as one of the Bi lattice constants.

The switches of particle-hole asymmetry in Figs. S8 and S9 are unlikely intrinsic, because they happen even for pairs with the same configuration but at different locations on Bi films. They are more likely from variation in the local environment between the different pairs. In these cases, for extracting the YSR states splitting energy, we use $||\Omega_2| - |\Omega_1||$ instead of $|\Omega_2 - \Omega_1|$, because both $\Omega_1$ and $\Omega_2$ have the same sign in most cases.

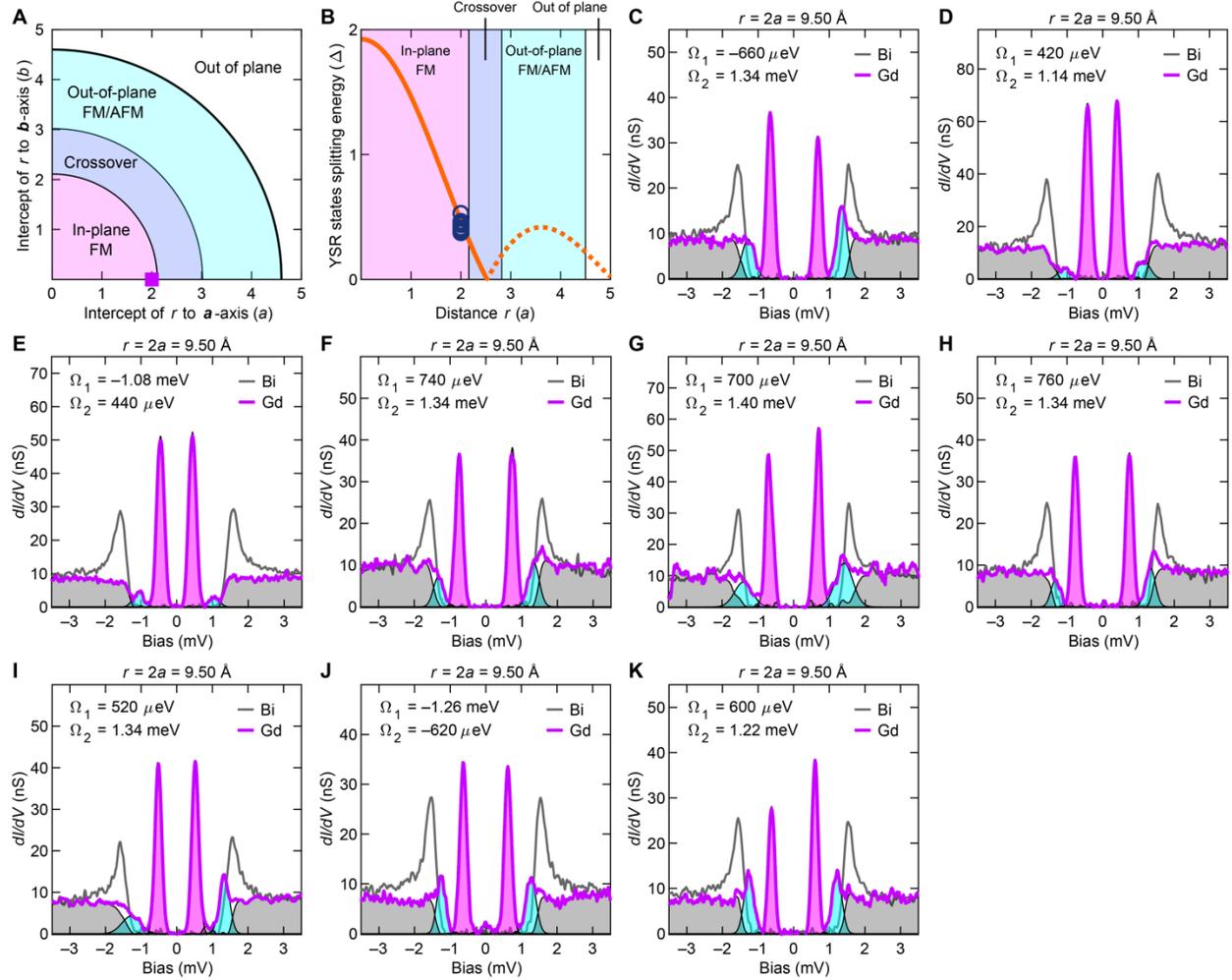

**Fig. S9.** YSR states of Gd pairs with separation $r = 2a$. (**A**) Its configuration in the phase diagram as in Fig. 1I. (**B**) YSR states splitting energies in the phase diagram as in Fig. 2C. (**C**–**K**) $dI/dV$ spectra taken on Gd pairs with separation $r = 2a$ at different locations on Bi films. $a = 4.75$ Å, as one of the Bi lattice constants.

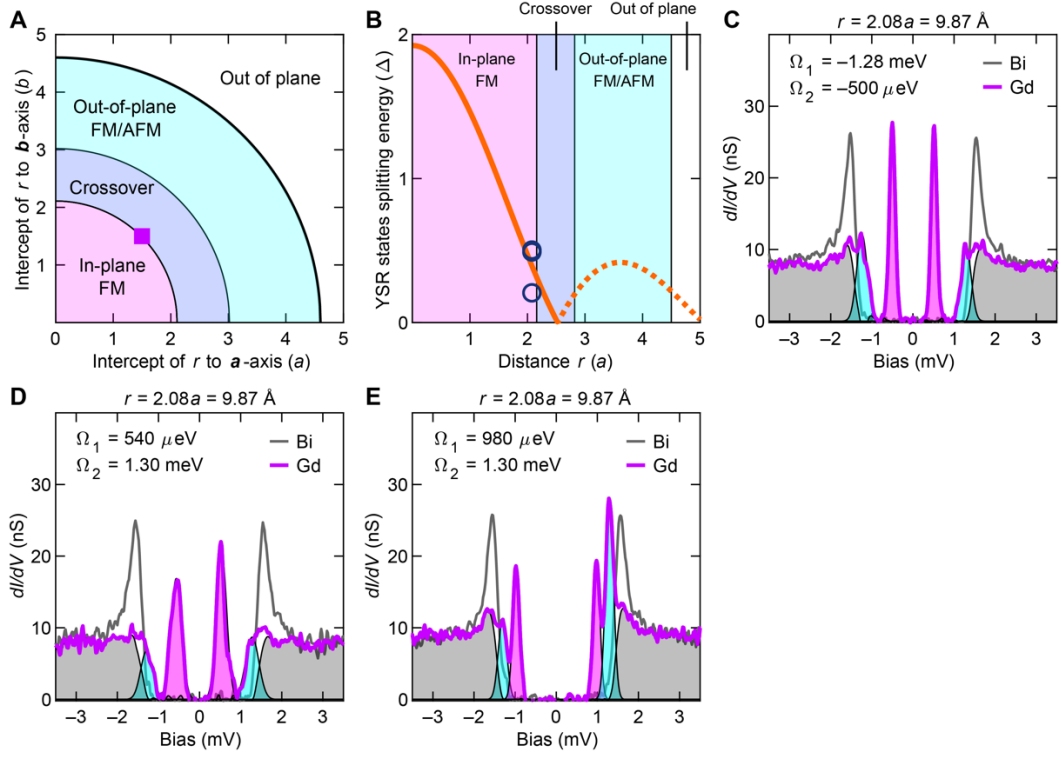

**Fig. S10.** YSR states of Gd pairs with separation $r = 2.08a$. (**A**) Its configuration in the phase diagram as in Fig. 1I. (**B**) YSR states splitting energies in the phase diagram as in Fig. 2C. (**C**–**E**) $dI/dV$ spectra taken on Gd pairs with separation $r = 2.08a$ at different locations on Bi films. $a = 4.75$ Å, as one of the Bi lattice constants.

Section V. Antiferromagnetic YSR states without splitting

For Gd pairs with atomic separations $r = 2.45a$ (Fig. S11) and $2.54a$ (Fig. S12), YSR states very reproducibly exhibit only one pair of degenerate peaks with electron-like state energy ($\Omega$) around $0.8\Delta = 1.2$ meV, which indicates the Gd spins in the pair are AFM aligned.

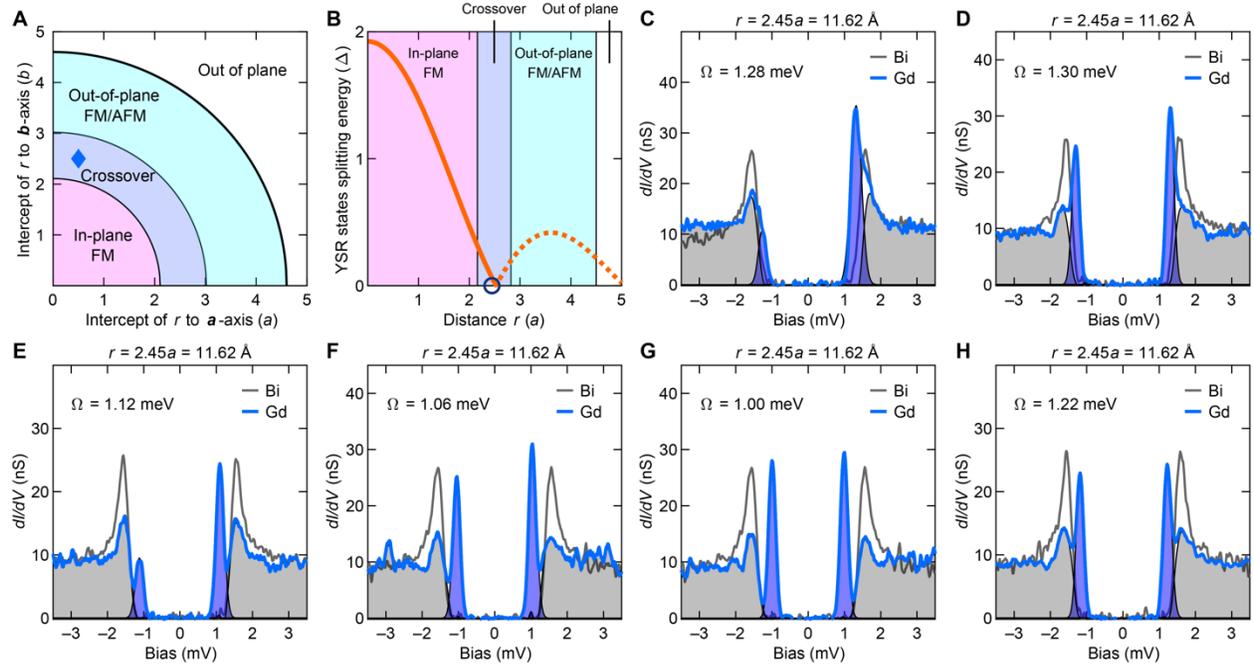

**Fig. S11.** YSR states of Gd pairs with separation $r = 2.45a$. (**A**) Its configuration in the phase diagram as in Fig. 1I. (**B**) YSR states splitting energies in the phase diagram as in Fig. 2C. (**C–H**) $dI/dV$ spectra taken on Gd pairs with separation $r = 2.45a$ at different locations on Bi films. $a = 4.75$ Å, as one of the Bi lattice constants.

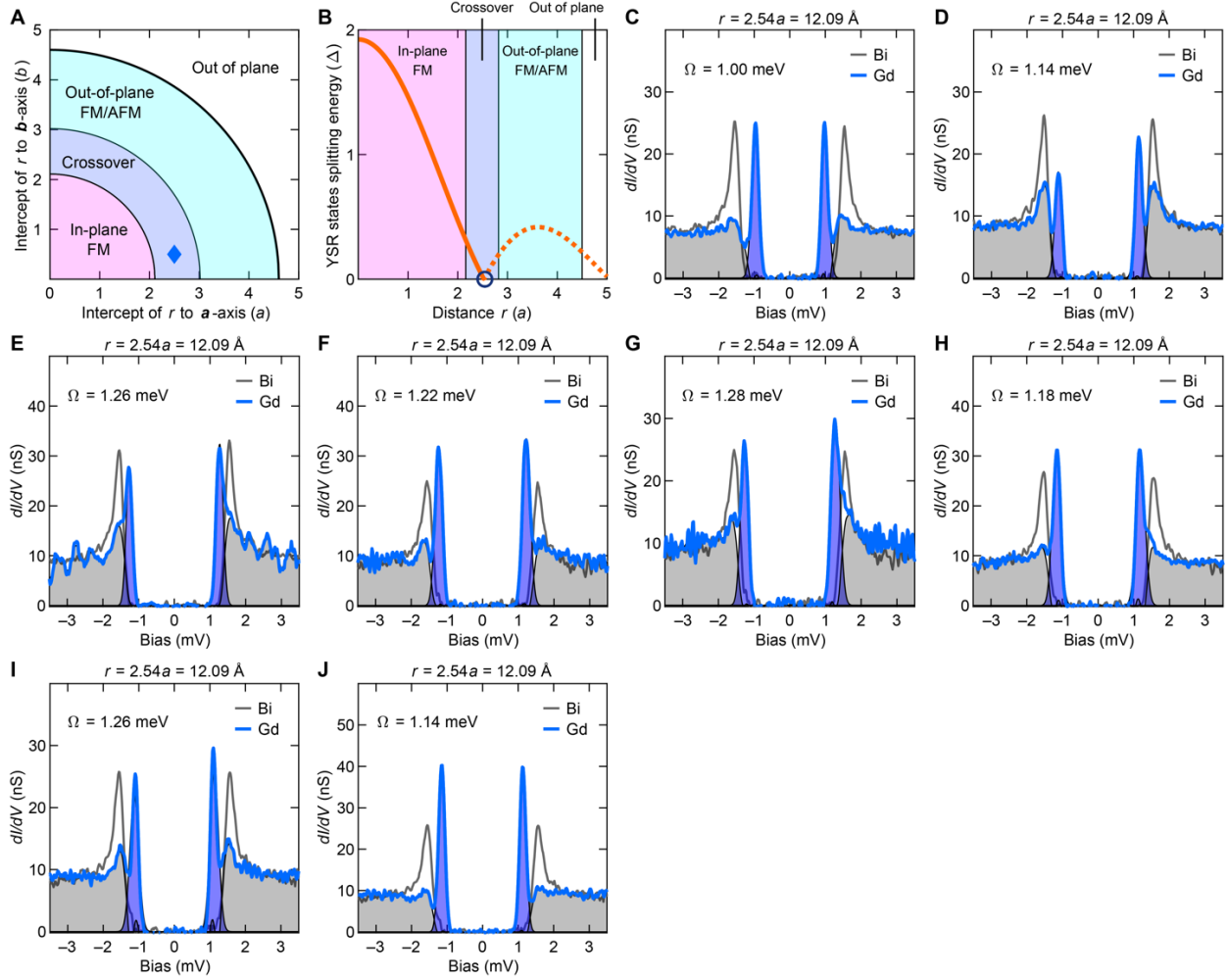

**Fig. S12**. YSR states of Gd pairs with separation $r = 2.54a$. (**A**) Its configuration in the phase diagram as in Fig. 1I. (**B**) YSR states splitting energies in the phase diagram as in Fig. 2C. (**C**–**J**) *dI/dV* spectra taken on Gd pairs with separation $r = 2.54a$ at different locations on Bi films. $a = 4.75$ Å, as one of the Bi lattice constants.

Section VI. YSR states with or without splitting

For Gd pairs with atomic separations $r = 2.16a$ (Fig. S13) and $2.21a$ (Fig. S14), which are in between FM regime ($r \leq 2.08a$) and AFM regime ($r \sim 2.5a$), YSR states show either 4 peaks (consistent with FM alignment) or 2 peaks (as expected for AFM alignment) from location to location on the surface. The spatial variability in spin arrangements occurs because the energy difference between FM and AFM states is very small in this crossover region, and a small local perturbation can make the spin ground state switch between them.

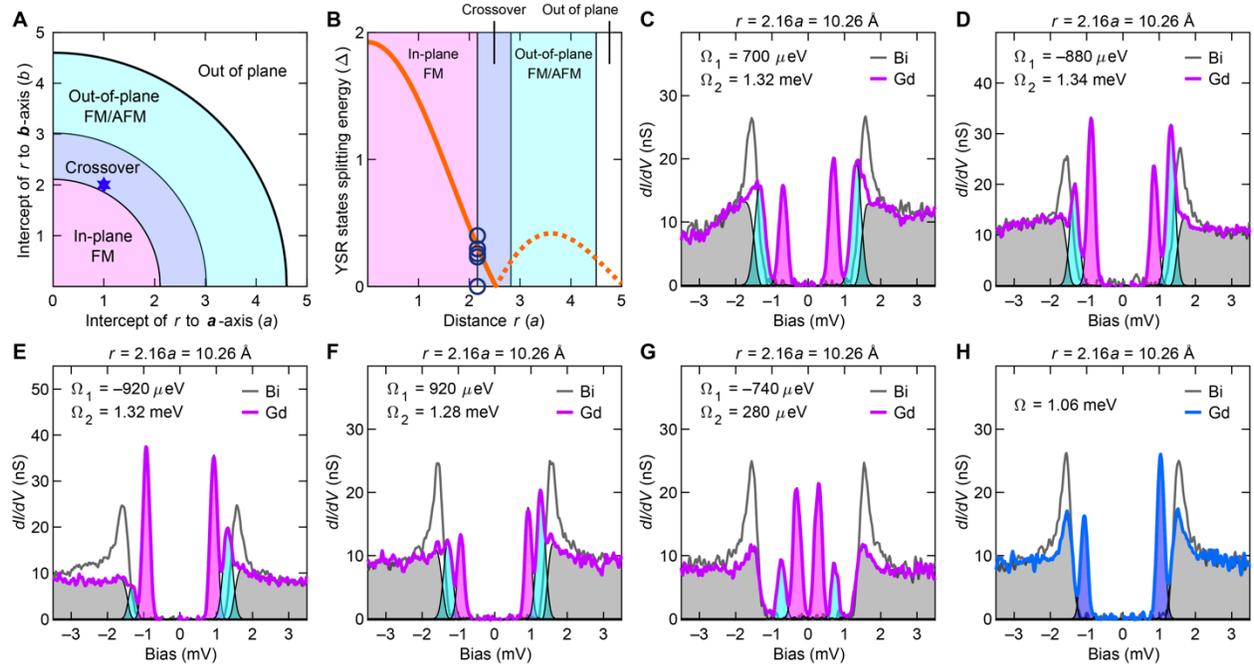

**Fig. S13.** YSR states of Gd pairs with separation $r = 2.16a$. (**A**) Its configuration in the phase diagram as in Fig. 1I. (**B**) YSR states splitting energies in the phase diagram as in Fig. 2C. (**C**–**H**) $dI/dV$ spectra taken on Gd pairs with separation $r = 2.16a$ at different locations on Bi films. $a = 4.75$ Å, as one of the Bi lattice constants.

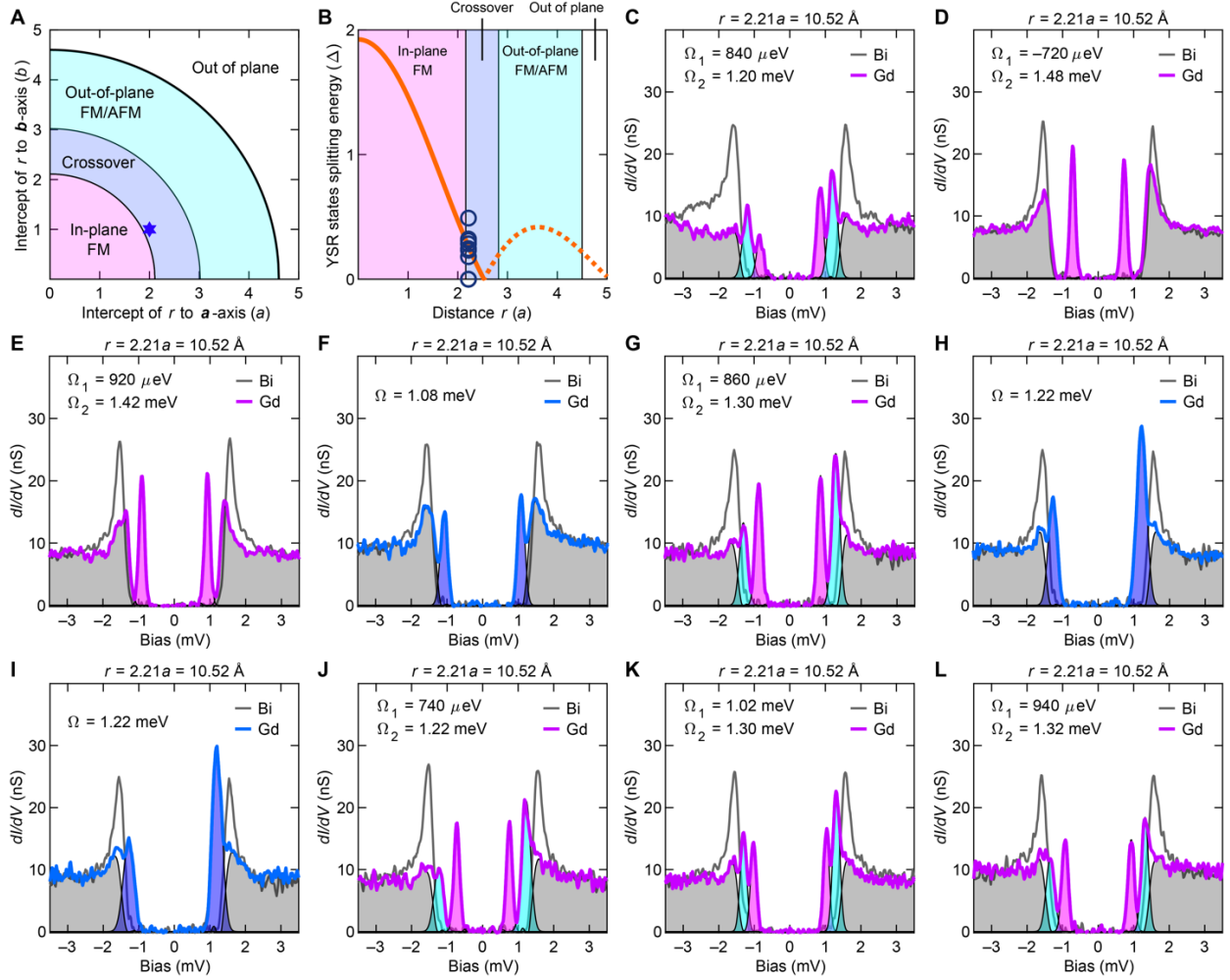

**Fig. S14.** YSR states of Gd pairs with separation $r = 2.21a$. (**A**) Its configuration in the phase diagram as in Fig. 1I. (**B**) YSR states splitting energies in the phase diagram as in Fig. 2C. (**C**–**L**) $dI/dV$ spectra taken on Gd pairs with separation $r = 2.21a$ at different locations on Bi films. $a = 4.75$ Å, as one of the Bi lattice constants.

For Gd pairs with atomic separations $r = 2.77a$ (Fig. S15) and $2.82a$ (Fig. S16), surface magnetic anisotropy competes with in-plane ordering and favors the Gd spins to point out of plane (see Section XI for details). In between in-plane FM/AFM regime ($r \leq 2.54a$) and out-of-plane FM/AFM regime ($r \geq 2.87a$), YSR states show either 4 peaks or 2 peaks with ($0.85\Delta < \Omega < \Delta$) depending on the location on the sample. In this regime, the energy difference between in-plane FM/AFM and out-of-plane FM/AFM states is very small in the transition region, and a small local perturbation can make the spin ground state switch between these ground states.

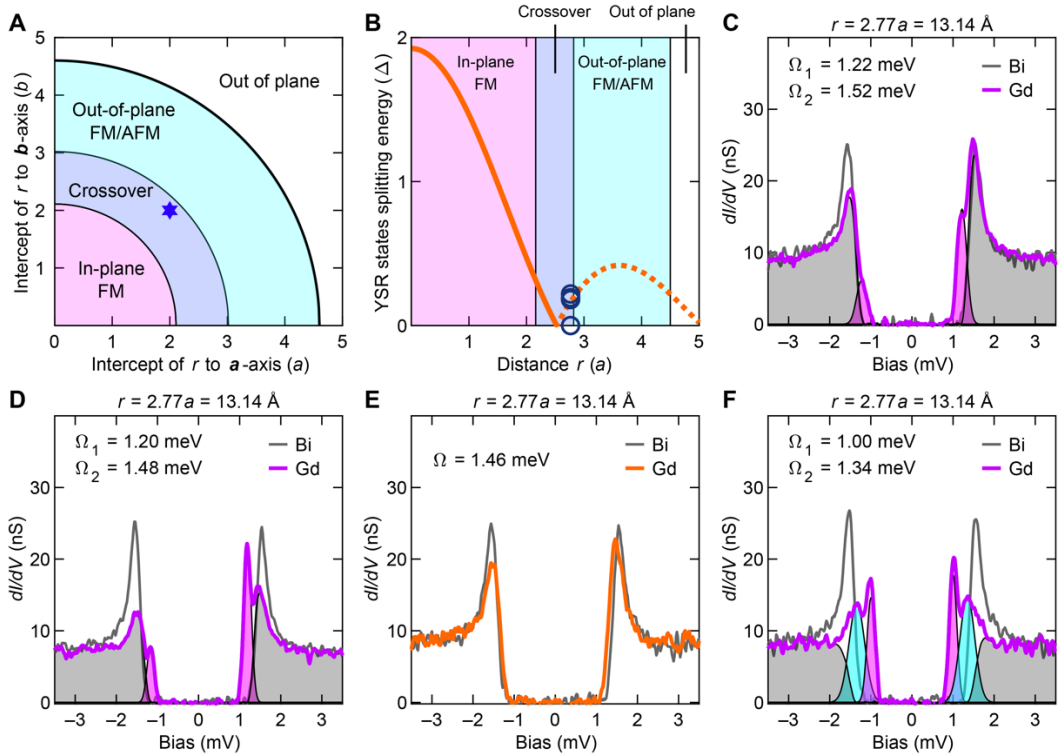

**Fig. S15.** YSR states of Gd pairs with separation $r = 2.77a$. (**A**) Its configuration in the phase diagram as in Fig. 1I. (**B**) YSR states splitting energies in the phase diagram as in Fig. 2C. (**C**–**F**) $dI/dV$ spectra taken on Gd pairs with separation $r = 2.77a$ at different locations on Bi films. $a = 4.75$ Å, as one of the Bi lattice constants.

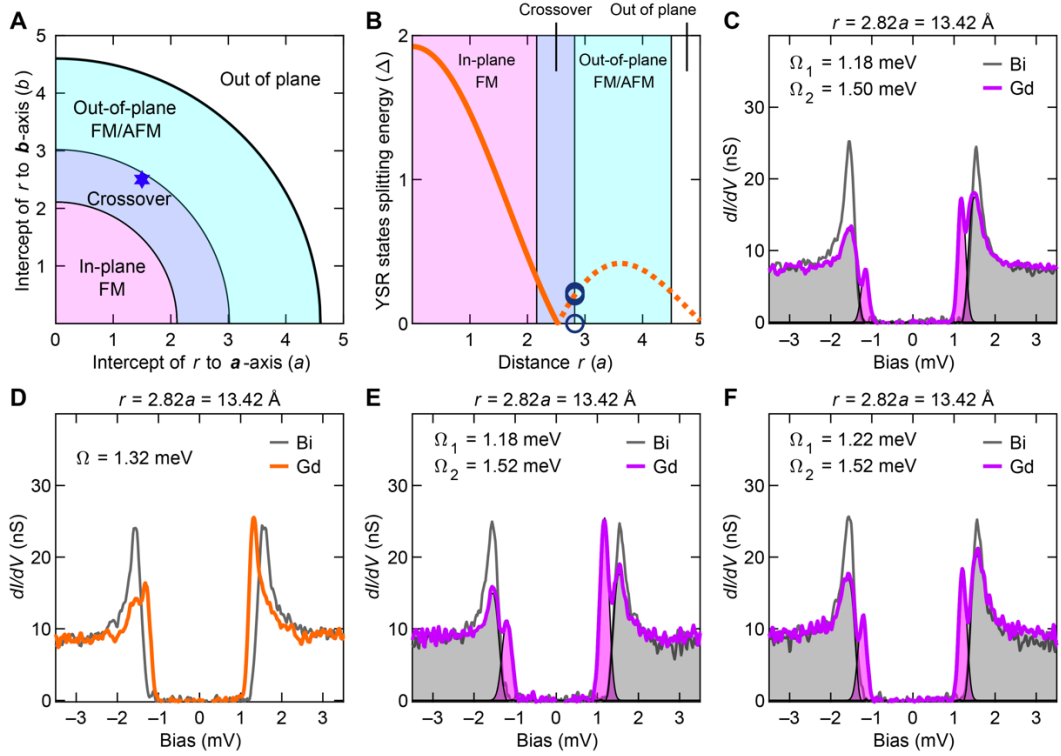

**Fig. S16.** YSR states of Gd pairs with separation $r = 2.82a$. (**A**) Its configuration in the phase diagram as in Fig. 1I. (**B**) YSR states splitting energies in the phase diagram as in Fig. 2C. (**C**–**F**) $dI/dV$ spectra taken on Gd pairs with separation $r = 2.82a$ at different locations on Bi films. $a$ = 4.75 Å, as one of the Bi lattice constants.

## Section VII. YSR states with $0.85\Delta < \Omega \leq \Delta$

For Gd pairs with atomic separations $r \geq 2.87a$ (Figs. S17–S20), surface magnetic anisotropy favors the Gd spins to point out of plane (see Section XI for details). In this situation only weak YSR states near the gap edge are observed.

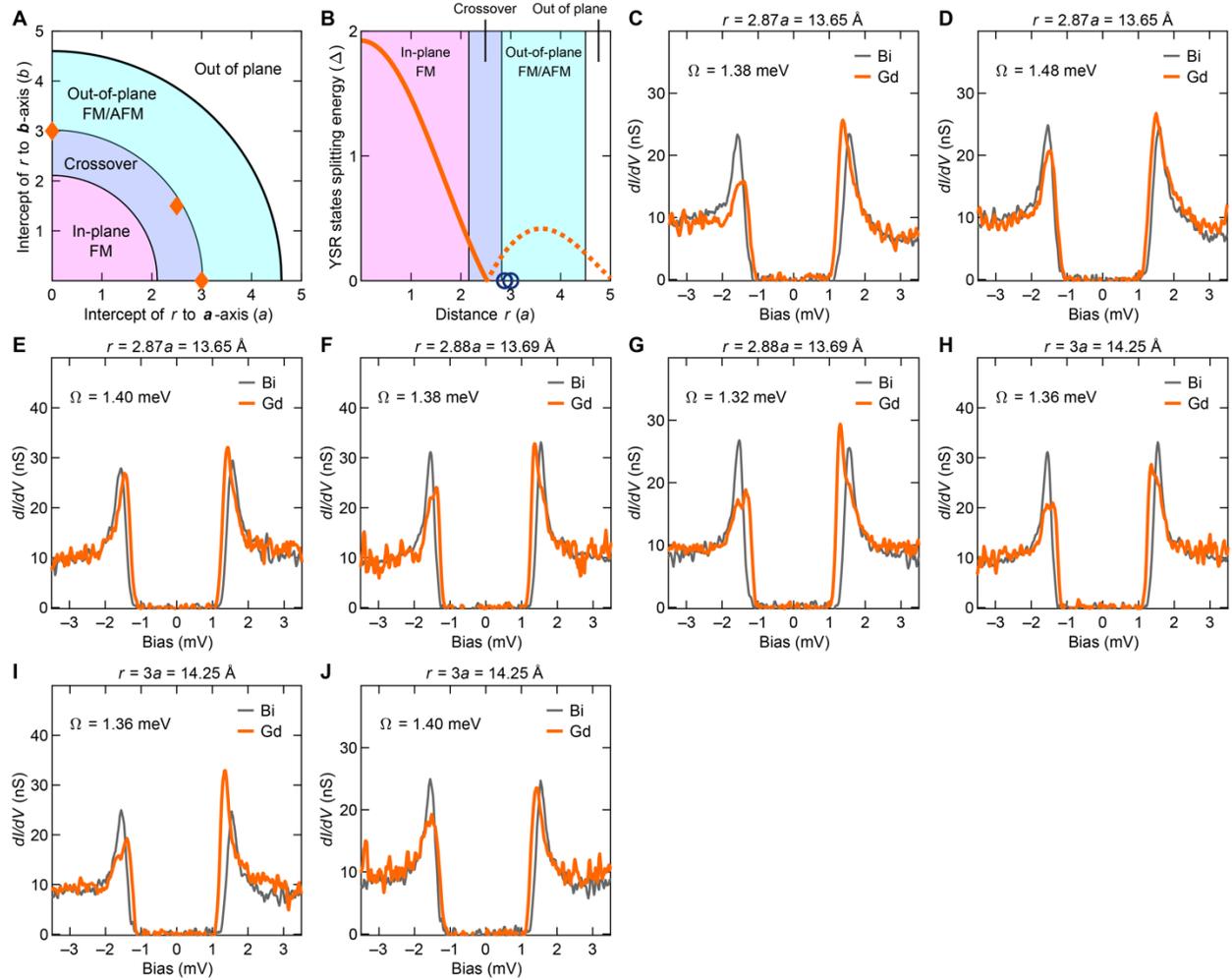

**Fig. S17.** YSR states of Gd pairs with separation $r = 2.87a$, $2.88a$, and $3a$. (**A**) Their configurations in the phase diagram as in Fig. 1I. (**B**) YSR states splitting energies in the phase diagram as in Fig. 2C. (**C**–**J**) $dI/dV$ spectra taken on Gd pairs with separation $r = 2.87a$, $2.88a$, and $3a$ at different locations on Bi films. $a = 4.75$ Å, as one of the Bi lattice constants.

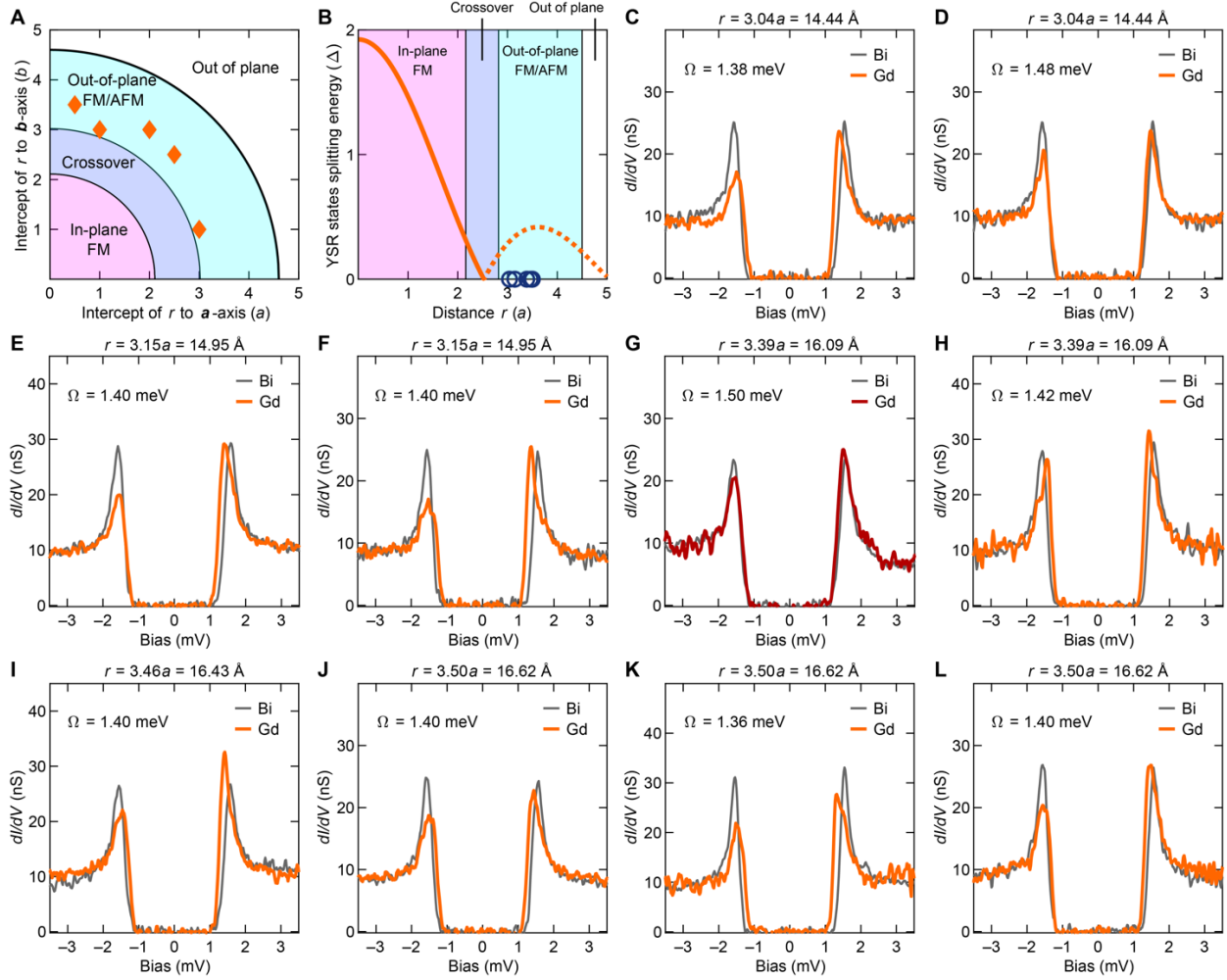

**Fig. S18.** YSR states of Gd pairs with separation $r = 3.04a$, $3.15a$, $3.39a$, $3.46a$, and $3.50a$. (**A**) Their configurations in the phase diagram as in Fig. 1I. (**B**) YSR states splitting energies in the phase diagram as in Fig. 2C. (**C**–**L**) $dI/dV$ spectra taken on Gd pairs with separation $r = 3.04a$, $3.15a$, $3.39a$, $3.46a$, and $3.50a$ at different locations on Bi films. $a = 4.75$ Å, as one of the Bi lattice constants.

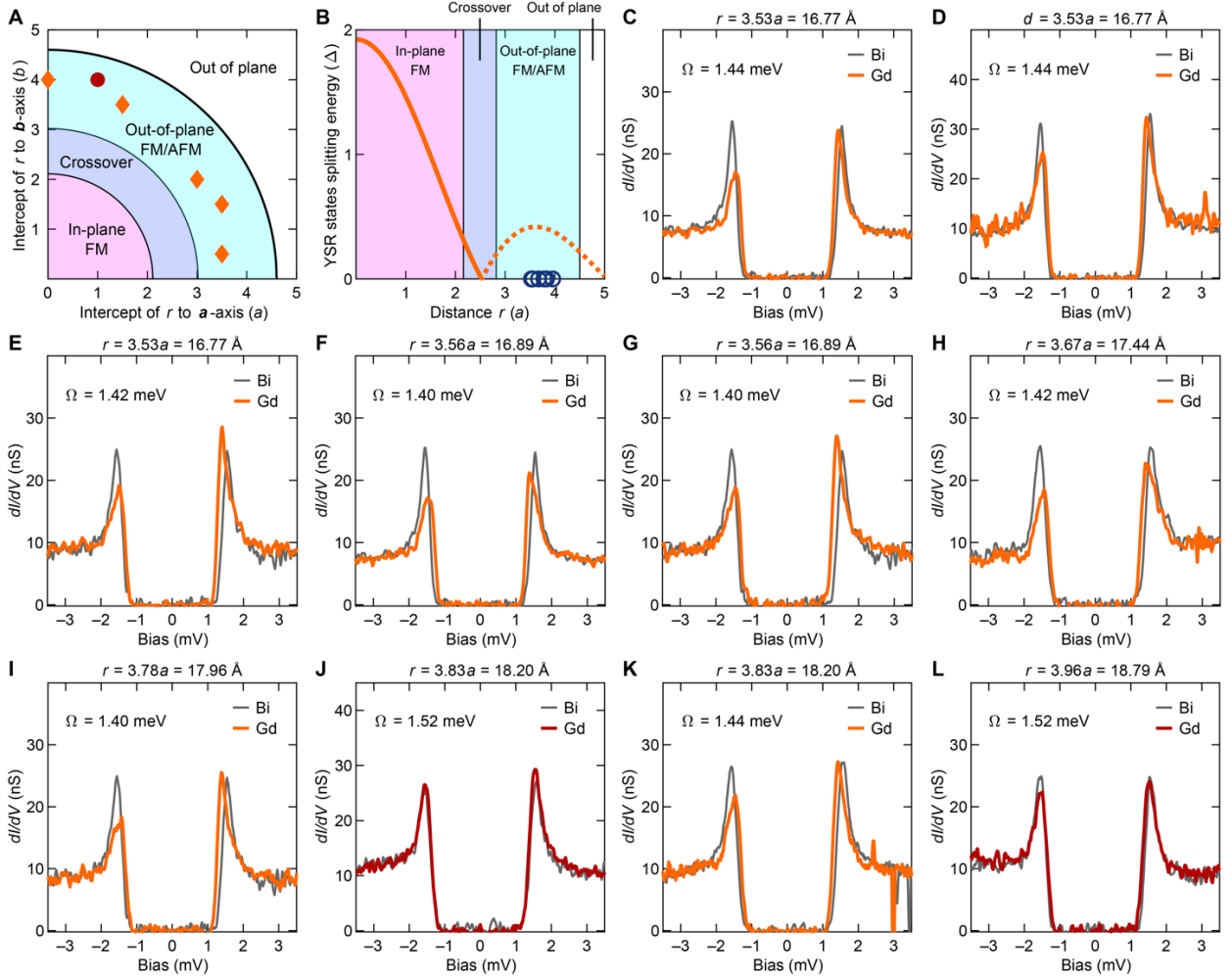

**Fig. S19.** YSR states of Gd pairs with separation $r = 3.53a$, $3.56a$, $3.67a$, $3.78a$, $3.83a$, and $3.96a$. (**A**) Their configurations in the phase diagram as in Fig. 1I. (**B**) YSR states splitting energies in the phase diagram as in Fig. 2C. (**C**–**L**) $dI/dV$ spectra taken on Gd pairs with separation $r = 3.53a$, $3.56a$, $3.67a$, $3.78a$, $3.83a$, and $3.96a$ at different locations on Bi films. $a = 4.75$ Å, as one of the Bi lattice constants.

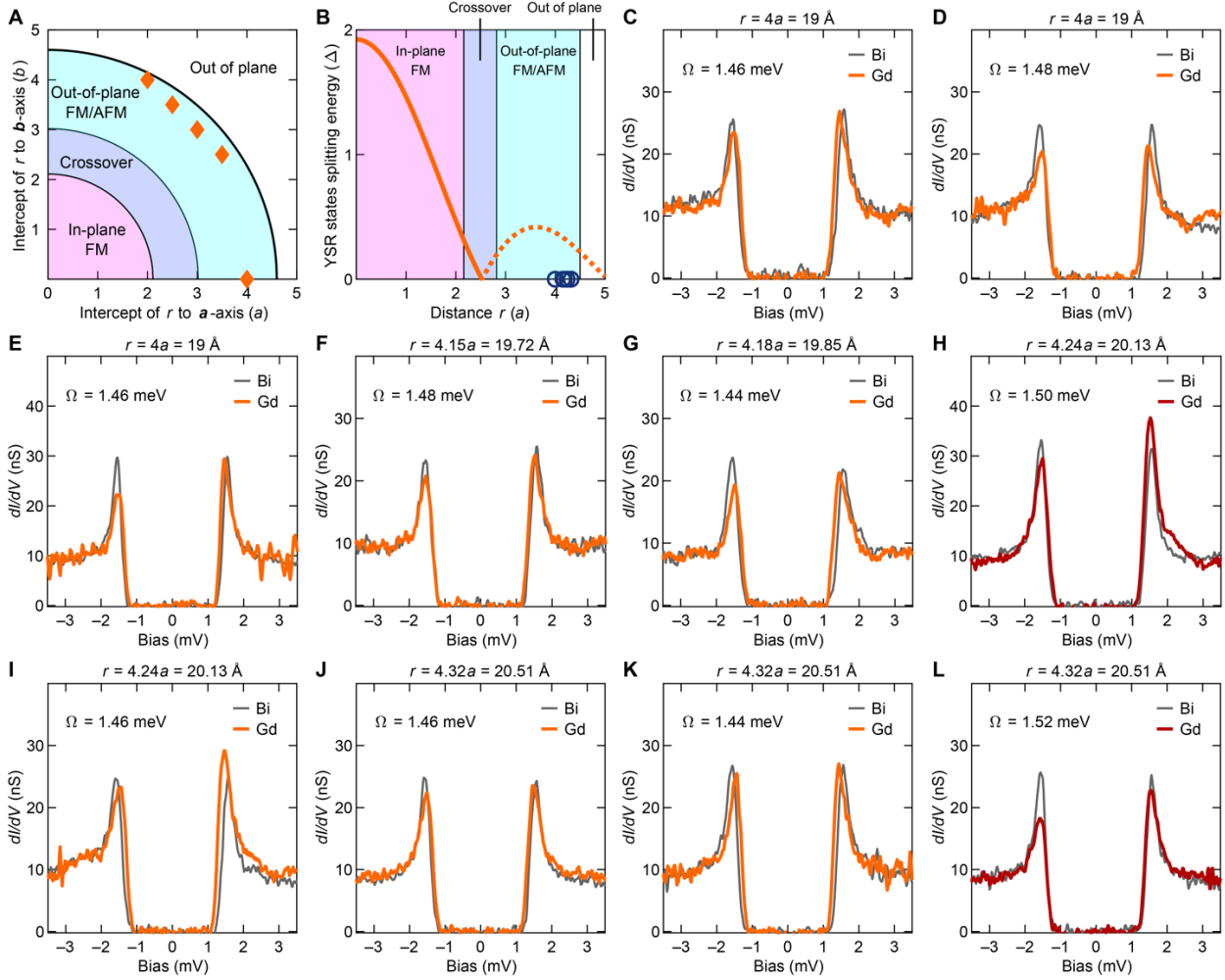

**Fig. S20.** YSR states of Gd pairs with separation $r = 4a$, $4.15a$, $4.18a$, $4.24a$, and $4.32a$. (**A**) Their configurations in the phase diagram as in Fig. 1I. (**B**) YSR states splitting energies in the phase diagram as in Fig. 2C. (**C**–**L**) $dI/dV$ spectra taken on Gd pairs with separation $r = 4a$, $4.15a$, $4.18a$, $4.24a$, and $4.32a$ at different locations on Bi films. $a = 4.75$ Å, as one of the Bi lattice constants.

## Section VIII. YSR states with $\Omega \sim \Delta$

For Gd pairs with atomic separations $r > 4.4a$ (Fig. S21), YSR states are identical to single Gd atom YSR states with $\Omega \sim \Delta$ (see Section XI for details).

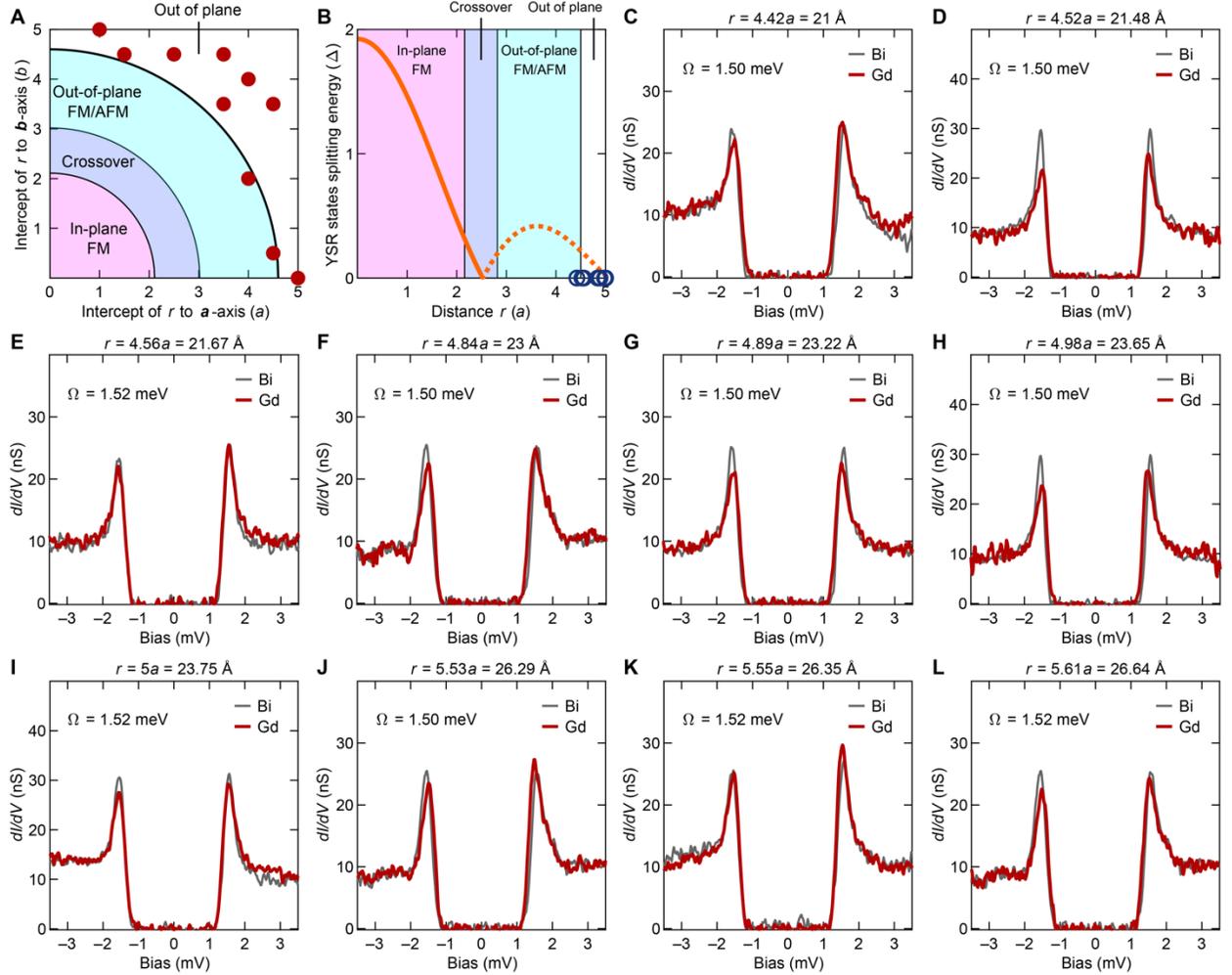

**Fig. S21.** YSR states of Gd pairs with separation $r = 4.42a$, $4.52a$, $4.56a$, $4.84a$, $4.89a$, $4.98a$, $5a$, $5.53a$, $5.55a$, and $5.61a$. (**A**) Their configurations in the phase diagram as in Fig. 1I. (**B**) YSR states splitting energies in the phase diagram as in Fig. 2C. (**C**–**L**) $dI/dV$ spectra taken on Gd pairs with separation $r = 4.42a$, $4.52a$, $4.56a$, $4.84a$, $4.89a$, $4.98a$, $5a$, $5.53a$, $5.55a$, and $5.61a$ at different locations on Bi films. $a = 4.75$ Å, as one of the Bi lattice constants.

Section IX. Gd pair without YSR states: a special case

In Fig. S22, we explore the YSR states for a pair of Gd atoms with very close interatomic distances where both RKKY interaction and direct exchange interaction are likely to co-exist. We start from a pair as shown in Fig. S22A with interatomic distance $r = 1.52a$. Its spectrum is shown in Fig. S22D where two pairs of YSR states located at –1.4 mV and 0 mV are observed. Then we move the bottom right atom in the pair towards upper left to make a $r = a$ pair as shown in Fig. S22B. After this manipulation, the spectrum of $r = a$ pair is nearly particle-hole symmetric and barely shows any YSR state (Fig. S22E). After this procedure, the pair moves as a unit as shown in Fig S22C, while still preserves its magnetic state with a very weak YSR state close to the gap edge (Fig. S22F). The manipulation sequence indicates that such a pair may in fact be a molecular dimer, different than pairs that we are focused on examining in this paper that interact through the superconductor. It is possible that such a dimer has a net zero spin because of AFM alignment due to direct exchange interaction and therefore does not couple to superconducting state.

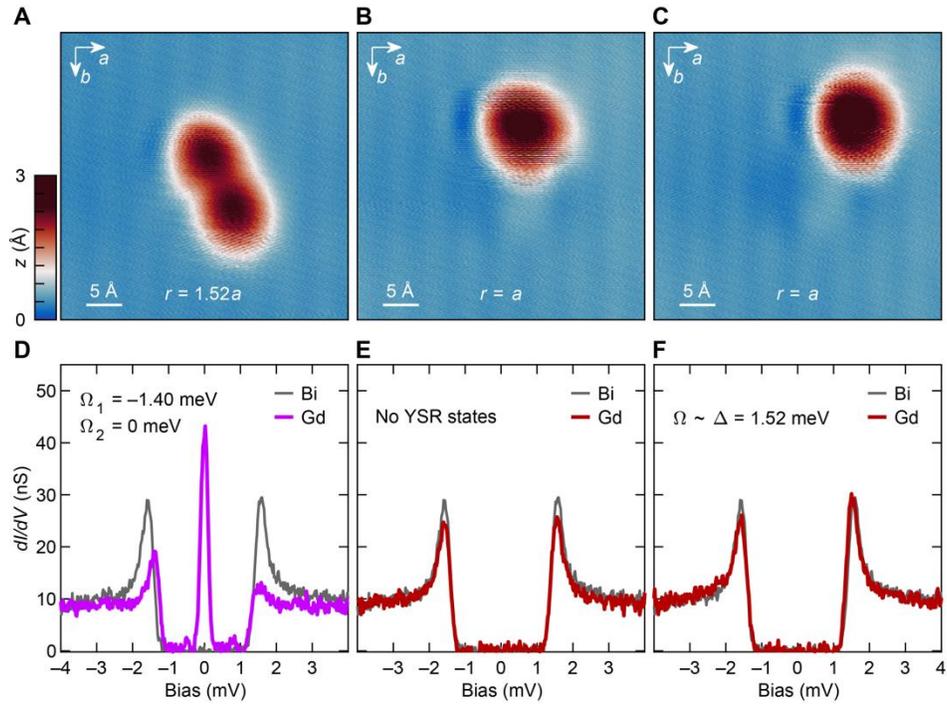

**Fig. S22.** A Gd pair without YSR states. (**A**–**C**) A manipulation sequence that moves the bottom right atom in panel A with $r = 1.52a$ towards upper left to make the pair with $r = a$ in panel B, and then move the pair as a whole unit in panel C. (**D**–**F**) $dI/dV$ spectra measured on the pairs in panels A–C, respectively.

Section X. YSR states of single Gd atoms measured with a superconducting tip

To distinguish the YSR states induced by a single Gd atom from the superconducting gap edge, we use a superconducting tip to increase the energy resolution. By crashing the tip into the superconducting Nb substrate, we coat the W tip with Nb atoms and make it superconducting. The superconductivity of the tip is calibrated by taking *dI/dV* spectrum on Bi surface (gray curve in Fig. S23), which shows a gap with $\Delta = \Delta_{Bi} + \Delta_{tip} = 2.62$ meV. Therefore, the tip exhibits a superconducting gap $\Delta_{tip} = (2.62 - 1.52)$ meV $= 1.10$ meV. With such a tip, we are able to resolve the single Gd atom YSR states from the coherence peaks. The electron-like state with higher spectral intensity is at energy $\Omega = 1.40$ meV (red curve in Fig. S23).

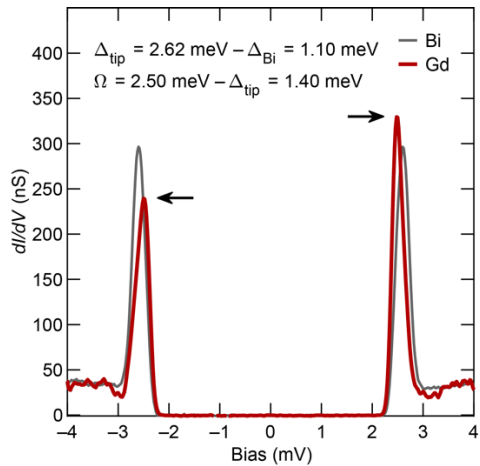

**Fig. S23.** YSR states of a single Gd atom measured with a superconducting tip. Gray: *dI/dV* spectrum measured on Bi surface away from Gd atoms with a superconducting tip. Red: *dI/dV* spectrum measured on a single Gd atom with the same tip.

Section XI. Theoretical model

To model the interactions between spins as a function of their separation and to determine its influence on their YSR states, we utilize the effect of anisotropy in a 2-dimensional electron gas (2DEG) system with large Rashba type spin-orbit coupling. In particular, we consider an anisotropic exchange interaction that is much larger in the plane of the 2DEG, the *xy*-plane, than along the axis perpendicular to it, the *z*-axis, competing with an on-site anisotropy which favors the spins to be along the *z*-axis.

Here we explicitly write down the exchange interaction between the impurity and the quasi-particles in the 2DEG as

$$H_\mathrm{I} = -\frac{1}{2}\int \Psi^\dagger(\boldsymbol{r}) \sum_{a,j} J_a S_j^a \sigma^a \delta(\boldsymbol{r}-\boldsymbol{r}_j)\Psi(\boldsymbol{r})d^2\boldsymbol{r}, \qquad (\text{S1})$$

where $a = x, y, z$ and $j = 1, 2$ labels the impurities. Here, $\boldsymbol{r}_j$ and $S_j^a$ are the position and spin of impurity $j$ and $\Psi(\boldsymbol{r})$ annihilates an electron at $\boldsymbol{r}$. We assume the exchange interaction between electrons and the magnetic impurity is anisotropic and $0 < J_z \ll J_x = J_y$. This is a consequence of both the anisotropy of the impurity atomic orbitals and the surface electron spin polarization (in-plane due to strong Rashba type spin-orbit coupling in Bi), giving rise to exchange couplings $J_a$ that are different along different directions. In addition, there is also the on-site anisotropy term

$$H_\mathrm{A} = -A\sum_j \left(S_j^z\right)^2, \qquad (\text{S2})$$

where the anisotropy constant $A > 0$ for a Gd atom with 7/2 spin. This anisotropy is caused by atomic spin-orbit interaction and broken symmetry of the atomic orbitals on the surface.

We first consider the case of an isolated magnetic impurity with spin $S$. There are two extreme conditions for the spin: (1) The spin points out of plane and gives rise to YSR states characterized by $\alpha_z = \pi S J_z \nu_\mathrm{F}$, where $S$ is the magnitude of the spin and $\nu_\mathrm{F}$ is the density of states of the underlying superconductor in the metallic phase at the chemical potential. Since $J_z$ is small we expect the YSR state to be in the weak coupling regime with $\alpha_z \ll 1$; in this case the YSR state lies at the gap edge with energy $\Omega = \frac{1-\alpha_z^2}{1+\alpha_z^2}\Delta \sim \Delta$. (2) The spin points in plane (without losing generality we assume the spin is along $x$-direction because the spin-orbit interaction is isotropic in the $xy$-plane). In this case the YSR states are characterized by $\alpha_x = \pi S J_x \nu_F$, which is of order of unity. Then the single atom should have a YSR state deep inside the gap at $\Omega = \frac{1-\alpha_x^2}{1+\alpha_x^2}\Delta$. From the Hamiltonian ($H_I + H_A$) we learn that, without exchange interaction between magnetic atoms and for a sufficiently large anisotropy, a single atom has a spin pointing out of plane dominated by the surface anisotropy term $H_A$. Thus, a single atom only has weak YSR states at gap edge, which is consistent with the experimental observation (for example, Figs. S3E and S23).

Next, we consider a second magnetic impurity at separation $r$ away from the first impurity along the $y$-direction. In the limit where $r$ is much smaller than the coherence length ($\xi$), the correction of the interaction from the superconducting electrons can be neglected and we get the RKKY interaction between the two magnetic impurities under perturbation with approximately the same form as in the metallic case in 2D (6, 7):

$$H_{\text{RKKY}} = -\frac{m}{2\pi^2\hbar^2}\frac{\sin(2q_F r)}{r^2}\left[J_x^2 S_1^x S_2^x + \left(J_y^2 S_1^y S_2^y + J_z^2 S_1^z S_2^z\right)\cos\theta_R \right. \\ \left. + J_y J_z \left(S_1^y S_2^z - S_1^z S_2^y\right)\sin\theta_R\right], \quad (S3)$$

where $q_F = \sqrt{2mE_F/\hbar^2 + k_{SO}^2}$, $E_F$ is the Fermi energy, $k_{SO}$ is the spin-orbit wave-vector, and $\theta_R = 2k_{SO} r$ corresponding to the twisting in the exchange interaction arising from the non-zero spin-orbit term. As $J_x, J_y \gg J_z$, the RKKY interaction leads to spin ground states either parallel or anti-parallel along the $x$-direction in the absence of the on-site anisotropy. However, for non-zero on-site anisotropy, there is a competition between the on-site anisotropy and RKKY terms; the former prefers spins to point along the *z*-axis. Because the RKKY interaction scales as the inverse square of separation *r* between the impurities, the magnetic ground state will depend on *r*. Quantitatively, for a sufficiently close separation *r* where the RKKY interaction overcomes the surface magnetic anisotropy, the spin ground state will be along *x*-direction. For sufficiently large *A* and *r*, the magnetic impurities are oriented along the *z*-axis and we expect the YSR states to be near the gap edge for large *r* as $J_z$ is small. With decreasing *r*, the RKKY interaction strength grows and becomes larger than the on-site anisotropy below a critical separation, favoring the pair of spins to align collinearly along the *x*-axis. At this critical separation we expect a jump in the energy of YSR states from the gap edge to the middle of gap as $J_x \gg J_z$. Furthermore, as $J_x$ is relatively large, the hybridization between the two nearby YSR states is more pronounced compared to the out-of-plane case. This is consistent with the experimental observation that the YSR state energies shift deep inside the gap at small separations and get split (for example, Figs. 1 E–H, 2 A and B).

Analytically, we can find the ground state configuration by minimizing the free energy *F* = $H_A$ + $H_{RKKY}$. Note that because the twisting of the spins in the RKKY Hamiltonian is only between the *y*- and *z*-components, we can focus on spins in the *yz*-plane. Upon finding the minimum energy spin configuration in the *yz*-plane, we compare that energy to $-|F(r)| = -\left|\frac{m}{2\pi^2\hbar^2}\frac{\sin(2q_F r)}{r^2}\right|$, which is the energy when the magnetic impurities spins are along the *x*-axis. The minimum of these two quantities determines the magnetic ground state.

Parameterizing $\mathbf{S}_j = (0, \sin\theta_j, \cos\theta_j)$ for spins twisting in the *yz*-plane, we find that either $\theta_1 = -\theta_2$ or $\theta_1 = \theta_2 - \pi$ due to the symmetry, and we have

$$\tan 2\theta_2 = -\frac{2F(r)J_y J_z \sin\theta_R}{2A + F(r)(J_y^2 + J_z^2)\cos\theta_R} \tag{S4}$$

or

$$\tan 2\theta_2 = \frac{2F(r)J_y J_z \sin\theta_R}{2A - F(r)(J_y^2 + J_z^2)\cos\theta_R}, \tag{S5}$$

respectively. For $J_z = 0$, $\theta_2 = 0, \pi/2$, which means $\theta_1 = 0, -\pi/2$ or $\theta_1 = \pi, \pi/2$. These correspond to the FM and AFM configurations collinear with either the $y$- or $z$-axis. Therefore, for small $J_z$, we expect the canting to be minimal if $|2A \pm F(r)J_y^2 \cos\theta_R| \gg 2|F(r)J_z^2 \cos\theta_R|$. Conceivably there could be some canting if $2A \pm F(r)(J_y^2 + J_z^2)\cos\theta_R = 0$, however the parameter space is very limited. And the energy for spins along $y$-direction is not as low as for the spins aligned along $x$-direction. Then the most energetically favorable ground states are spins aligned collinearly either along $x$- or $z$-directions.

To validate the above scenario with numerical analysis, in which the RKKY interaction is calculated as a function of separation, we consider a band structure with a Dirac point being away from the Fermi level and has Rashba type spin-orbit interaction (similar to that in Ref. 6) described by the tight-binding Hamiltonian

$$H = H_0 + H_{SC} + H_{SOI} + H_I, \tag{S6}$$

where

$$H_0 = -t\sum_{\langle i,j\rangle,\sigma=\uparrow,\downarrow}(c_{i,\sigma}^\dagger c_{j,\sigma}) - \mu\sum_{i,\sigma}(c_{i,\sigma}^\dagger c_{i,\sigma}), \tag{S7}$$

$$H_{SC} = \Delta\sum_i (c_{i,\uparrow}^\dagger c_{i,\downarrow}^\dagger - c_{i,\downarrow}^\dagger c_{i,\uparrow}^\dagger + \text{H.c.}), \tag{S8}$$

$$H_{SOI} = \lambda\sum_i\left[(c_{i+\hat{x},\downarrow}^\dagger c_{i,\uparrow} - c_{i+\hat{x},\uparrow}^\dagger c_{i,\downarrow}) + i(c_{i+\hat{y},\downarrow}^\dagger c_{i,\uparrow} + c_{i+\hat{y},\uparrow}^\dagger c_{i,\downarrow}) + \text{H.c.}\right], \tag{S9}$$

$$H_{\mathrm{I}} = \sum_{\substack{\sigma,\sigma'=\uparrow,\downarrow \\ a=x,y,z}} \left(J_a c_{\mathbf{n},\sigma}^{\dagger} \sigma_{\sigma,\sigma'}^{a} c_{\mathbf{n},\sigma} + J_a c_{\mathbf{m},\sigma}^{\dagger} \sigma_{\sigma,\sigma'}^{a} c_{\mathbf{m},\sigma}\right), \tag{S10}$$

where $c_{\mathbf{i},\sigma}$ is the annihilation operator of an electron with spin $\sigma$ on a square lattice at site $\mathbf{i} = (i_x, i_y)$. The sum over $\mathbf{i}$ runs over all sites of the $N_x \times N_y$ lattice and the sum $\langle \mathbf{i}, \mathbf{j} \rangle$ runs over neighboring sites. The parameters $t$, $\mu$, $\Delta$, and $\lambda$ are the hopping amplitude, chemical potential, superconducting pairing, and spin-orbit interaction strength, respectively. The impurities reside at lattice points $\mathbf{n}$ and $\mathbf{m}$ and are anisotropically exchange coupled to the electrons with magnitude $J_a$ along the $a$-axis, where $a = x, y, z$.

For an initial set of tight-binding parameters on a $32 \times 32$ lattice, we take $t = 1$, $\mu = 4$, $\lambda_x = \lambda_y = 0.5$, $\Delta = 0.1$, and $J_x = J_y = 10 J_z = 1.7$ corresponding to the nearest neighbor hopping, chemical potential, spin-orbit interaction, proximity-induced superconducting gap, and anisotropic exchange, respectively. This system has Fermi wavelength $\lambda_F = 10a$, where $a$ is the lattice constant. For these parameters, we find that a single YSR state is at $\Omega = 0.45\Delta$ when the impurity spin is oriented along the $x$-axis, or at the gap edge when the impurity spin is along the $z$-axis.

To determine the magnetic ground state, we find the difference between the energy of the magnetic impurity along the $x$-axis: $E_x$, and the energy of the magnetic impurity along the $z$-axis: $E_z$. We assume $J_x = J_y = 10 J_z = J$ for $\Delta = 0.1$, and $\Delta = 0.05$. The anisotropy must be greater than the difference in the energies of these two configurations, specifically $A > E_z - E_x$, for the single impurity spin to prefer to point along the $z$-axis. When $J = 1.7$ for instance, the critical anisotropy for which the magnetic ground state is collinear with the $z$-axis is $A_c = 0.07$.

Using the same parameters, we compare the energies of two magnetic impurities configured collinearly along the $x$-axis and configured collinearly along the $z$-axis. In Fig. S24 we plot the energy difference between the impurity configurations collinear with respect to each other along the $x$-axis and aligned parallel to the $z$-axis as a function of separation between them.

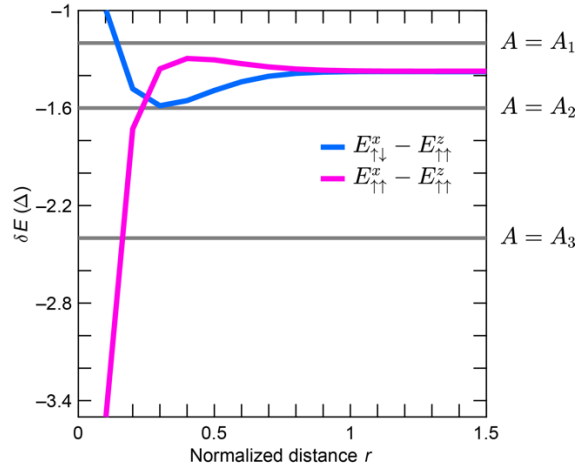

**Fig. S24.** Spin ground states of Gd pairs determined by the competition between the RKKY interaction and the on-site anisotropy. Magenta (blue) curve is the energy difference of the RKKY term between the in-plane FM (AFM) state and the out-of-plane FM state. Gray lines denote the energy difference of the on-site anisotropic term between the in-plane and out-of-plane spin configurations, with $A_1$, $A_2$, $A_3$ denoting three degrees of anisotropy strength in the system. At different separations, the spin ground state corresponds to the lowest energy state among these two curves and the gray lines.

For a small anisotropy, for instance in this case when $A = A_1 = 0.6\Delta$, the magnetic configuration is dominated by the RKKY interaction. When magnetic impurities are next to each other, the magnetic ground state is in the parallel FM configuration (magenta curve) but switches to the antiparallel AFM configuration (blue curve) as the separation between them is increased. Plotting the YSR states energies in Fig. S25A, we see they strongly split in the parallel magnetic configuration when $r < 0.3$ and are degenerate when $r \geq 0.4$. If the anisotropy is not too large or small, e.g., $A = A_2 = 0.8\Delta$, we expect the magnetic ground state to be along the $z$-axis when $r \geq 0.4$, and to be along the $x$-axis otherwise (Fig. S25B). The relative alignment of spins in plane depends on the sign of the RKKY interaction. Because $J_z \ll J_x$, the YSR states shift to the gap edge when $r \geq 0.4$. For a large anisotropy, $A = A_3 = 1.2\Delta$, the RKKY interaction is strong enough to overcome the anisotropy only when the impurities are very close to each other; otherwise, the YSR states remain at the gap edge for $r \geq 0.2$ (Fig. S25C).

The anisotropy we have chosen is simply a phenomenological parameter used to match the results of our model calculations with the experimental measurements of the YSR energies. Assuming $A = A_2$, we find that the behavior of the numerically calculated YSR energies as a function of $r$ agree with the experimental measurements (Fig. S25B).

Here we notice that in the experiment, the in-plane to out-of-plane spin ground state transition happens at a larger $k_F r$ value compared to the calculation. This is because the experimental system is between 2D and 3D, while the calculation is based on a 2D model where spatial overlap of the YSR states has been underestimated at distances $r < \lambda_F$.

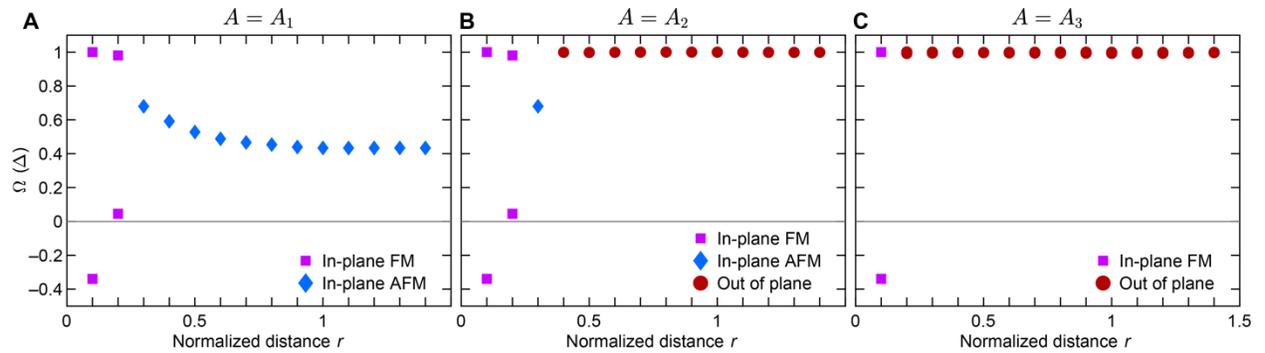

**Fig. S25.** Interatomic distance dependence of YSR states energies of Gd pairs at different degrees of anisotropy. (**A**–**C**) YSR states energies as a function of distance under the anisotropy strength $A = A_1, A_2, A_3$, respectively.

Section XII. Degeneracy of YSR states in the presence of spin-orbit coupling

Now that we identify the spin ground state of a pair of Gd atoms with $k_F r \sim 1$ is either in-plane FM or in-plane AFM depending on the sign of RKKY interaction, we then investigate whether the degeneracy of YSR states of Gd pairs can reflect the spin orientation. From theory we know that in a system without spin-orbit coupling, a pair of FM aligned spins in a superconductor creates a pair of bonding and anti-bonding YSR states with energy splitting proportional to wavefunction overlap of the YSR states localized around the two spins. But for a pair of AFM aligned spins, the YSR states are degenerate, because their wavefunctions are orthogonal and therefore the overlap is strictly zero.

However, this is usually not the case in systems with spin-orbit coupling (8). With the inclusion of the in-plane Rashba type spin-orbit coupling, YSR states from a pair of AFM aligned spins are no longer degenerate because the spin-orbit coupling can bring non-zero wavefunction overlap between the YSR states from the two spins. In a system with both spin-orbit coupling and RKKY interaction, we rewrite Hamiltonian in Eqn. S3 as:

$$H'_{\text{RKKY}} = -\frac{m}{2\pi^2 \hbar^2} \frac{\sin(2 q_F r)}{r^2} \big[ S_1'^x S_2'^x + \big( S_1'^y S_2'^y + S_1'^z S_2'^z \big) \cos \theta_R \\ + \big( S_1'^y S_2'^z - S_1'^z S_2'^y \big) \sin \theta_R \big],$$  (S11)

where $S_j'^a = J_a S_j^a$, $a = x, y, z$ and $j = 1, 2$ labels the impurities. Under a coordinate transformation:

$$\begin{cases} S_j''^x = S_j'^x \\ S_j''^y = S_j'^y \cos \theta_R + S_j'^z \sin \theta_R, \\ S_j''^z = S_j'^z \cos \theta_R - S_j'^y \sin \theta_R \end{cases}$$  (S12)

the Hamiltonian becomes:

$$H''_{\text{RKKY}} = -\frac{m}{2\pi^2\hbar^2}\frac{\sin(2q_F r)}{r^2}\left(S_1''^x S_2''^x + S_1''^y S_2''^y + S_1''^z S_2''^z\right), \tag{S13}$$

in the rotated frame.

Therefore, the spin ground state in the rotated frame is either FM or AFM, depending on the relative amplitude of $S_j''^a$, where $a = x, y, z$ and $j = 1, 2$. For the FM or AFM spin alignment along *y*- or *z*-direction in the rotated frame, it corresponds to a non-collinear spin alignment in the real coordinates, according to Eqns. S11–S13. The AFM spin ground state in the rotated frame shows degenerate YSR states under the effect of spin-orbit coupling, whereas the AFM state in the real coordinates has partially FM components in the rotated frame, thus shows non-degenerate YSR states. But for the FM or AFM spin alignment along the *x*-direction in the rotated frame, the spins are still collinear along the *x*-direction in the real coordinates (Eqns. S11–S13). As a consequence, the YSR states induced by AFM aligned spins along the *x*-direction remain degenerate even with the inclusion of in-plane Rashba type spin-orbit coupling. Therefore, the fact that we have degenerate YSR states localized around a pair of Gd atoms with separation $r \sim 2.5a$ is consistent with our theoretical analysis that the degenerate YSR states reflect in-plane AFM ground state, with the spins aligned perpendicularly to the connection between two atoms.

Section XIII. Extracting the couplings and non-collinear spin alignment in 3-atom chains

When spin pairs are FM aligned, the finite overlap between their YSR states break the degeneracy and split them into a pair of bonding and anti-bonding states. Assuming the isolated single spin YSR state has an energy $\Omega_0$ and the coupling matrix element between the two YSR states is $t$, then the bonding and anti-bonding states appear at energies $\Omega_{B/AB} = \Omega_0 \pm t$.

We can extend this model to the 3-atom chain, which provides important insights into how spin alignment evolves as we build longer chains. For FM spin alignment in such a chain, we denote the nearest neighbor coupling matrix element to be $t_1$, and the coupling matrix element between the next-nearest neighbor atoms (Atom 1 and Atom 3) to be $t_2$. Then the three energies of the hybridized YSR states are $\Omega_1 = \Omega_0 - t_2$, $\Omega_{2,3} = (2\Omega_0 + t_2 \pm \sqrt{8t_1^2 + t_2^2})/2$, corresponding to the non-bonding state, bonding state and anti-bonding state, respectively, in the 3-atom chain case. Using this model to analyze the additional splitting of YSR peaks measured on the two 3-atom chains with spacing $r = 2a$ in Fig. 3, we can extract the coupling parameters assuming the spins are FM aligned (Table S1).

|  | $\Omega_0$ (meV) | $t_1$ (meV) | $t_2$ (meV) |
| --- | --- | --- | --- |
| Gd pair, $r = 2a$, 4 peaks (Fig. 3A) | 0.90 | 0.30 |  |
| 3-Gd-atom chain, $r = 2a$, 6 peaks (Fig. 3D) | 0.81 | 0.30 | –0.04 |
| 3-Gd-atom chain, $r = 2a$, 6 peaks (Fig. 3I) | 0.84 | 0.27 | 0.10 |

**Table S1.** Coupling parameters extracted from the pair and 3-atom chain shown in Fig. 3, assuming all the spins are FM aligned.

Examining the results of this analysis, we find that the extracted next-nearest neighbor coupling matrix element $t_2$ is much smaller than $t_1$ in both 3-atom chains. This is in contradiction with the theoretical model (9) where $t_2$ is supposed to be comparable to $t_1$ ($t_2 \sim -0.8t_1$) for a FM aligned 3-atom chain with spacing $r = 2a$, since the energy splitting of the YSR states vary as $\left|\frac{\sin(k_F r)}{cr}\right|$ for FM aligned pairs as shown in Fig. 2C (orange curve), where $c = \frac{k_F(1+\alpha_x^2)}{4\alpha_x \Delta}$.

Therefore, the spin alignment of the 3-atom chain is unlikely FM, and that the observation of very small $t_2$ is the consequence of the non-collinear alignment of the spins. Theoretical study shows if there is an angle θ between the spins of a pair of magnetic atoms, the splitting between the YSR energies (*t*) will be reduced from the FM alignment (θ = 0°) by a factor of cos(θ/2) (Ref. 4). Therefore, the reduced YSR energy splitting in the 3-atom chain here could be resulted from a non-collinear alignment of the three spins.

Section XIV. Long-range spin-spin interactions between a 3-Gd-atom chain and a Gd atom

In Fig. 4, we show results of experiments in which we moved the fourth atom towards the 3-atom chain and examine the split YSR states on 3-atom chain following the same analysis in Section XIII. Here we simply treat the fourth atom as a perturbation to the 3-atom chain which slightly changes the coupling parameters $\Omega_0$, $t_1$, and $t_2$ of the chain (Table S2). As we move in the fourth atom, we find that the single atom YSR state energy $\Omega_0$ increasing and the nearest neighbor coupling matrix element $t_1$ decreasing monotonically. This implies that the angle between the nearest spins slightly increases when the fourth atom is brought closer to the 3-atom chain. Although a complete understanding of the spin alignment in such situation requires a detailed theory, our experiment demonstrates that long-range spin-spin interactions and non-collinear spin arrangements can be tuned in such a superconducting state.

|  | $\Omega_0$ (meV) | $t_1$ (meV) | $t_2$ (meV) |
|---|---|---|---|
| 3-Gd-atom chain ($r = 2a$) (Fig. S26A) | 0.81 | 0.30 | –0.04 |
| 3-Gd-atom chain ($r = 2a$) + Gd atom ($r_{34} = 3.39a$) (Fig. S26E) | 0.87 | 0.27 | –0.07 |
| 3-Gd-atom chain ($r = 2a$) + Gd atom ($r_{34} = 3.04a$) (Fig. S26J) | 0.88 | 0.24 | –0.02 |
| 3-Gd-atom chain ($r = 2a$) + Gd atom ($r_{34} = 2.16a$) (Fig. S26O) | 0.91 | 0.22 | –0.07 |

**Table S2.** Coupling parameters extracted from the 3-atom chain in Figs. 4 and S26, assuming all the spins of the 3-atom chain are FM aligned. $r_{34}$ is the distance between Atom 3 and Atom 4.

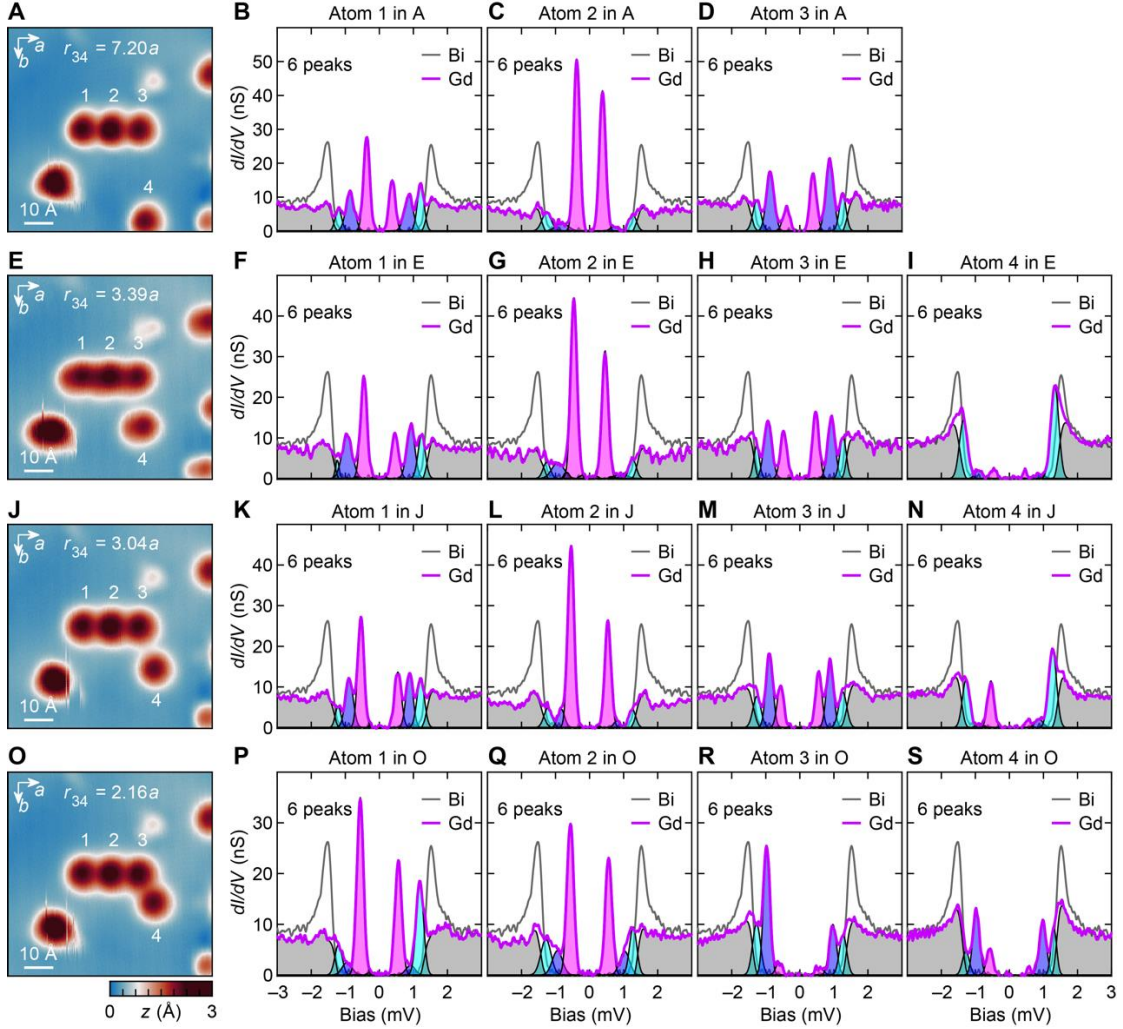

**Fig. S26.** Raw data of Fig. 4. (**A**–**D**) Topography and *dI/dV* spectra of a 3-atom chain evenly spaced by $r = 2a$ between nearest neighbor atoms. $r_{34} = 7.20a$ is the distance between Atom 3 and Atom 4. $a = 4.75$ Å, as one of the Bi lattice constants. (**E**–**I**) Topography and *dI/dV* spectra of the 3-atom chain and a fourth atom (Atom 4) which is moved closer towards the chain to $r_{34} = 3.39a$. (**J**–**N**) Topography and *dI/dV* spectra of the 3-atom chain and Atom 4 which is moved closer towards the chain to $r_{34} = 3.04a$. (**O**–**S**) Topography and *dI/dV* spectra of the 3-atom chain and Atom 4 which is moved closer towards the chain to $r_{34} = 2.16a$.

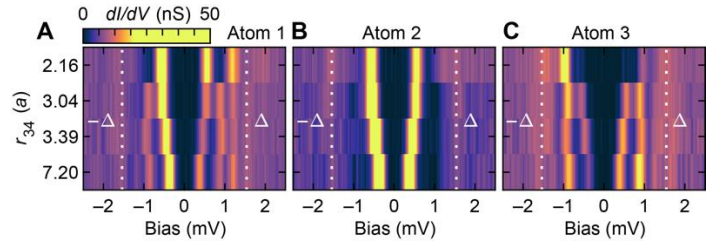

**Fig. S27.** Unprocessed data of Fig. 4 E–G. (**A**–**C**) Color plots of *dI/dV* spectra taken on Atoms 1 to 3 in panels A–D of Fig. 4, from bottom to top, showing YSR states evolution through the manipulation sequence. $r_{34}$ is the distance between Atom 3 and Atom 4. $a$ = 4.75 Å, as one of the Bi lattice constants.

**SI References**